\documentclass[
preprint,
 amsmath,amssymb,floatfix
 aps,
prm,
]{revtex4-2}

\usepackage{graphicx}
\usepackage{makecell,booktabs}
\usepackage{dcolumn}
\usepackage{bm}
\usepackage{hyperref}
\hypersetup{
    colorlinks=true,
    urlcolor= blue,
    citecolor=blue,
linkcolor= blue}
\usepackage{xcolor}

\begin{document}

\title{Constraints on proximity-induced ferromagnetism in a Dirac semimetal \texorpdfstring {(Cd$_3$As$_2$)}(/ferromagnetic semiconductor \texorpdfstring {(Ga$_{1-x}$Mn$_x$Sb)}(  heterostructure}
\author{Arpita Mitra}
\author{Run Xiao}
\author{Wilson Yanez}
\author{Yongxi Ou}
\affiliation{Department of Physics, The Pennsylvania State University, University Park PA 16803}
\author{Juan Chamorro}
\affiliation{Department of Chemistry, Johns Hopkins University, Baltimore, MD USA}
 \author{Alexander J. Grutter}
\affiliation{NIST Center for Neutron Research, National Institute of Standards and Technology, Gaithersburg, Maryland 20899, USA}
\author{Michael R. Fitzsimmons}
\affiliation{Spallation Neutron Source, Oak Ridge National Laboratory, Oak Ridge, Tennessee 37830, USA}
\author{Timothy R. Charlton}
\affiliation{Spallation Neutron Source, Oak Ridge National Laboratory, Oak Ridge, Tennessee 37830, USA}
\author{Tyrel McQueen}
\affiliation{Department of Chemistry, Johns Hopkins University, Baltimore, MD USA}
\author{Julie A. Borchers}
\affiliation{NIST Center for Neutron Research, National Institute of Standards and Technology, Gaithersburg, Maryland 20899, USA}
\author{Nitin Samarth}
\email{nsamarth@psu.edu}
\affiliation{Department of Physics, The Pennsylvania State University, University Park, Pennsylvania 16802, USA}

\date{\today}

\begin{abstract}
Breaking time-reversal symmetry in a Dirac semimetal Cd$_3$As$_2$ through doping with magnetic ions or by the magnetic proximity effect is expected to cause a transition to other topological phases (such as a Weyl semimetal). To this end, we investigate the possibility of proximity-induced ferromagnetic ordering in epitaxial Dirac semimetal (Cd$_3$As$_2$)/ferromagnetic semiconductor (Ga$_{1-x}$Mn$_x$Sb) heterostructures grown by molecular beam epitaxy. We report the comprehensive characterization of these heterostructures using structural probes (atomic force microscopy, x-ray diffraction, scanning transmission electron microscopy), angle-resolved photoemission spectroscopy, electrical magneto-transport, magnetometry, and polarized neutron reflectometry. Measurements of the magnetoresistance and Hall effect in the temperature range 2 K - 20 K show signatures that could be consistent with either a proximity effect or spin-dependent scattering of charge carriers in the Cd$_3$As$_2$ channel. 
Polarized neutron reflectometry sets constraints on the interpretation of the magnetotransport studies by showing that (at least for temperatures above 6 K) any induced magnetization in the Cd$_3$As$_2$ itself must be relatively small ($<$ 14 emu/cm$^3$). 

\end{abstract}
\maketitle

\section{\label{sec:level1}Introduction}
The study of the magnetic proximity effect at the interface between a ferromagnet and a nonmagnetic material has a long history. Early measurements explored the proximity effect in bilayers that interfaced magnetic metals (such as Cr, Fe, Ni) or ferromagnetic insulators (EuS, YIG) with a conventional nonmagnetic metal (Al, Pd, Pt) \cite{Hauser_PR_1969,Huang_PhysRevLett.109.107204}. More recently, the magnetic proximity effect has attracted attention within the context of topological quantum materials since it may provide a means of breaking time reversal symmetry (TRS) without the disorder introduced by magnetic doping. Most of these latter studies have focused on topological insulators such as Bi$_2$Se$_3$ \cite{Vobornik_WOS:000295667000009,JSLee_npjQM, Tang_WOS:000406370700063, Katmis_WOS:000376443100036,Riddiford_PhysRevLett.128.126802}, but the use of a magnetic proximity effect to perturb the symmetry-protected states can be extended to all classes of topological quantum materials. Dirac semimetals (DSMs) provide an interesting opportunity in this context. In a DSM, the stability of the Dirac node depends on three important symmetries: crystal, inversion, and TRS \cite{Crassee}. Breaking TRS in a DSM via a magnetic field is expected to result in a transition to a Weyl semimetal \cite{Cano_PhysRevB.95.161306,Baidya_PhysRevB.102.165115}. Introducing magnetic dopants that lead to long-ranged ferromagnetic order within a DSM could accomplish the same result, but solubility of magnetic dopants in the lattice of a host DSM can present challenges \cite{Xiao_PhysRevMaterials.6.024203}. In this work, we seek an alternative pathway for breaking TRS in the states of the archetypal DSM Cd$_3$As$_2$ via proximity exchange coupling with a ferromagnetic semiconductor. 

Cd$_3$As$_2$ has attracted attention as a canonical DSM with good chemical stability \cite{Liu2014}, extraordinarily high carrier mobility (higher than $10^6 \mathrm{cm}^2 \cdot {\mathrm{(V.s)}}^{-1}$ at cryogenic temperatures in bulk crystals) \cite{Liang2015, Neupane2014}, and topologically nontrivial bulk bands created by a strong spin-orbit interaction \cite{Wang2013,Yang2014,Wehling2014}. This topological quantum material provides an excellent platform for realizing novel quantum phenomena such as the chiral anomaly and surface Fermi arcs. In recent years, much of the interest in this DSM has turned to the study of Cd$_3$As$_2$ thin films \cite{schumann2016molecular} where the interplay between quantum confinement and topology creates a rich playground for studying quantum transport \cite{uchida2017quantum,schumann2018observation,Galletti_PhysRevB.99.201401,Xiao_PhysRevB.106.L201101}, engineering different topological phases \cite{Shoron_2021}, and topological spintronics \cite{Yanez_PhysRevApplied.16.054031}. 

The DSM phase of Cd$_3$As$_2$ is tetragonal and belongs to the I4$_1$/acd centrosymmetric space group \cite{Cava} with lattice parameters a = 1.2633 nm and c = 2.5427 nm \cite{Steigmann}. The preferred growth direction for Cd$_3$As$_2$ thin films is $\langle112\rangle$. The bulk band structure of Cd$_3$As$_2$ is characterized by two Dirac nodes, the points where the conical conduction and valence bands meet, located along the tetragonal $z$~ axis of the crystal \cite{Wang2013}. These nodes are protected from the effects of disorder by \textit C$_\textit 4$ rotational symmetry \cite{Yang2014}. We use a ferromagnetic semiconductor rather than a metallic ferromagnet for trying to realize the magnetic proximity effect in order to minimize parallel transport in the FM layer; in principle, this allows transport studies of the impact of broken TRS on quantum transport in the DSM. We note that a recent study used transport measurements to deduce proximity-induced ferromagnetism in Cd$_3$As$_2$ grown on a highly insulating ferrimagnet (a rare earth iron garnet) using a combination of pulsed laser deposition and solid state epitaxy \cite{Uchida_PhysRevB.100.245148}. This is only possible using a post-growth, high temperature anneal of the samples after capping with a protective layer. Such an approach limits the possible heterostructure configurations and perhaps even the crystalline integrity. We also note that the direct doping of Cd$_3$As$_2$ thin films with magnetic atoms (transition metal or rare earth) is likely to be challenging due to phase separation and surfactant effects \cite{Xiao_PhysRevMaterials.6.024203}.

Since the magnetic proximity effect is typically mediated by a short-ranged exchange interaction \cite{Zutic_MT}, it is important to create a sharp, chemically-ordered interface between the ferromagnetic and non-magnetic materials. In this paper, we use molecular beam epitaxy (MBE) to synthesize hybrid DSM/ferromagnetic semiconductor heterostructures using materials that have compatible growth temperatures and compatible surface chemistry so that the heterointerface is well controlled.  Since the epitaxial growth by MBE of Cd$_3$As$_2$ films of reasonable quality on (111) GaSb/GaAs is well-established \cite{schumann2016molecular,Schumann2017,schumann2018observation,Galletti_PhysRevB.99.201401,Yanez_PhysRevApplied.16.054031, Xiao_PhysRevB.106.L201101}, we use Mn-doping of GaSb to create a ferromagnetic semiconductor ``substrate.'' Mn is an acceptor in GaSb, resulting in a p-doped ferromagnetic semiconductor at Mn concentrations greater than a few percent \cite{Matsukara_JAP}; by using a low enough Mn concentration ($\sim 3\%$), we can achieve Ga$_{1-x}$Mn$_x$Sb layers that host ferromagnetism while still remaining much more resistive than the overgrown Cd$_3$As$_2$ film. The use of a III-Mn-V ferromagnetic semiconductor to provide a magnetic interface with a topological quantum material is similar to our past demonstration of proximity-induced ferromagnetism in a canonical topological insulator \cite{JSLee_npjQM}. The only caveat is that, unlike better established III-Mn-V ferromagnetic semiconductors such as (Ga,Mn)As and (In,Mn)As, the ferromagnetism in (Ga,Mn)Sb has more complex origins due to a greater tendency for the formation of MnSb clusters \cite{Matsukara_JAP,MCCOMBE200390}.  At high substrate temperatures, Mn-doping of GaSb yields readily observable phase-separated ferromagnetic metal clusters of MnSb; additionally, the unusual surfactant behavior of Mn in the presence of Cd$_3$As$_2$ also creates some potential challenges. At low substrate temperatures, the incorporation of Mn is more uniform, but the ferromagnetic behavior of (Ga,Mn)Sb films in this regime still may involve inhomogeneous phases that are difficult to directly detect in microscopy \cite{Matsukara_JAP}.   

In this paper, we primarily study the magneto-transport properties of Cd$_3$As$_2$/Ga$_{1-x}$Mn$_x$Sb in the temperature range 2 K $\leq T \leq$ 20 K and show that the variation of the magnetoresistance (MR) and Hall effect with temperature and magnetic field indicates either a magnetic proximity effect or spin-dependent scattering at the Cd$_3$As$_2$/magnetic semiconductor interface when it is smooth and of high quality. 
For further confirmation of possible proximity effects, we then use polarized neutron reflectometry (PNR) to obtain the magnetization profile at the interface between the DSM and the ferromagnetic semiconductor. As shown recently, PNR measurements can be essential for ruling out misleading artifacts in transport measurements that can mimic a magnetic proximity effect \cite{PhysRevLett.128.126802}. Our PNR measurements allow us to set constraints on the interpretation of the magnetotransport, suggesting that at temperatures above 6 K, the signatures of magnetic coupling in the Cd$_3$As$_2$ layer are more likely due to spin-dependent scattering of charge carriers in the DSM rather than induced ferromagnetism. Our results are a first step toward the manipulation of the topological DSM phase via the magnetic proximity effect and identify the challenges that remain to achieve this goal. 

\begin{figure*}
\includegraphics[width=1\textwidth]{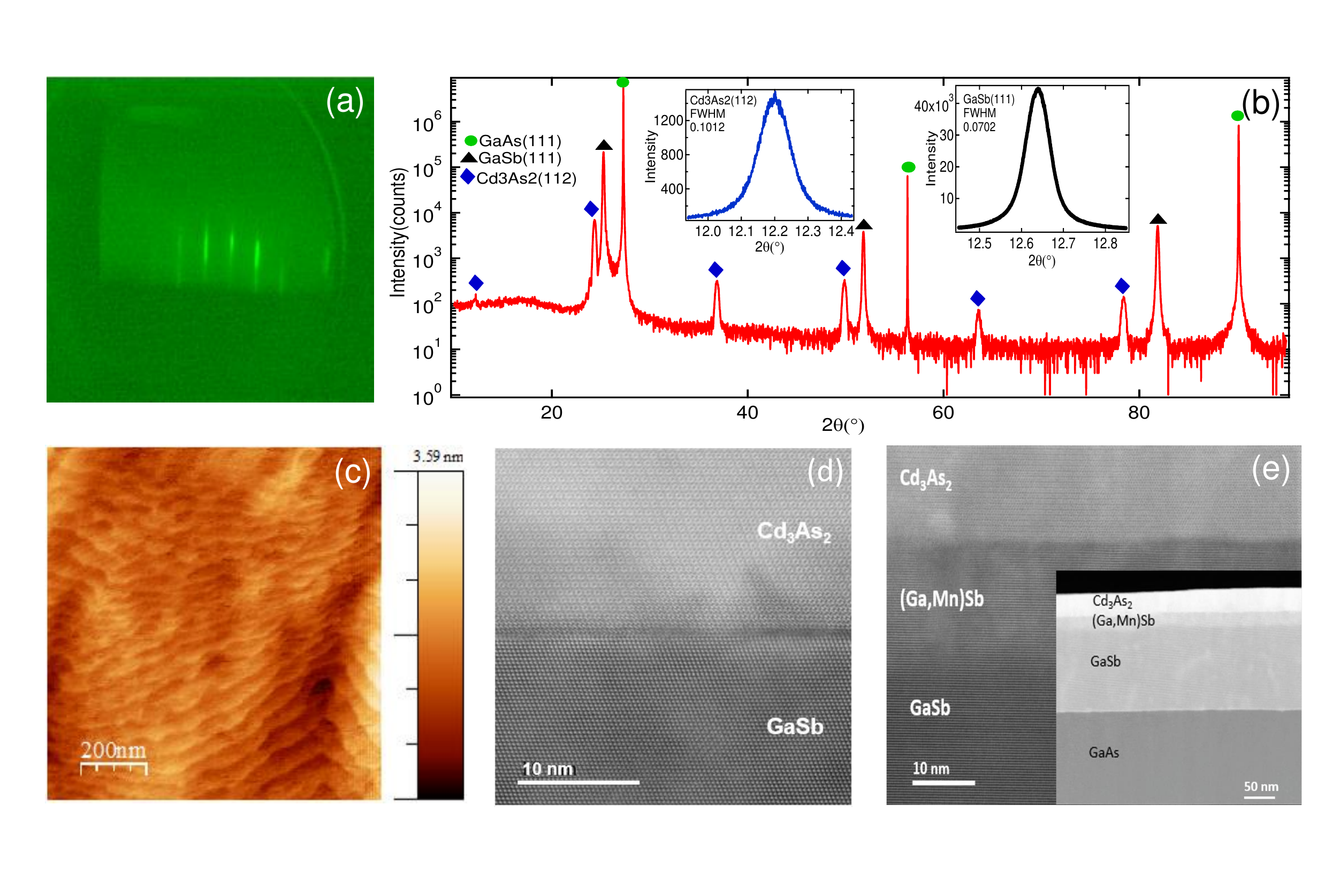}
\caption{\label{FIG.1} Structural characterization of type A MBE-grown Cd$_3$As$_2$/Ga$_{1-x}$Mn$_x$Sb heterostructures. (a) \textit {In situ} RHEED pattern taken along the $\langle 111 \rangle$ direction after deposition of Cd$_3$As$_2$ on Ga$_{1-x}$Mn$_x$Sb. (b) Out-of-plane XRD spectrum of a Cd$_3$As$_2$/Ga$_{1-x}$Mn$_x$Sb heterostructure. The marked peaks are from each material, showing Cd$_3$As$_2$ with a (112) sample surface plane, while other peaks come from the buffer layer and the substrate. Insets show the rocking curves of the GaSb buffer and Cd$_3$As$_2$, indicating good crystalline quality of the material. (c) AFM image showing smooth surface of Cd$_3$As$_2$ film with RMS roughness $\sim 1.2$ nm. HAADF-TEM images of (d) a Cd$_3$As$_2$/GaSb heterostructure and (e) a type A  Cd$_3$As$_2$/Ga$_{1-x}$Mn$_x$Sb heterostructure.}
\end{figure*}


\section{\label{sec:level2}Growth and Structural Characterization} 

 We carried out the growth of Ga$_{1-x}$Mn$_x$Sb/Cd$_3$As$_2$ heterostructures in a VEECO 930 MBE system 
 (base pressure lower than 2 x $10^{-10}$ mbar) using epi-ready GaAs (111) B miscut substrates tilted $\pm$ 1$^\circ$ toward the $[2 \bar{1} \bar{1}]$ direction. The purpose of using miscut substrates is to reduce the twinning in Cd$_3$As$_2$ thin films, which originates from the 5\% lattice mismatch. The entire growth process was monitored {\it in situ} by 12 keV reflection high-energy electron diffraction (RHEED). We used high purity (5N) elemental sources for Ga, As, and Mn, while for Cd$_3$As$_2$, we used a compound source prepared from high purity elemental sources. All materials were evaporated from standard effusion cells. After flashing the native oxide from the epi-ready GaAs substrates at $580^\circ \mathrm{C}$, we first grew the ferromagnetic semiconductor ``substrate" using two different approaches. In samples referred to as ``type A," the ferromagnetic substrate consists of a 60 nm thick GaSb buffer layer deposited at a substrate temperature $\sim$490$^\circ$C followed by a uniformly Mn-doped Ga$_{1-x}$Mn$_x$Sb (20 nm) layer grown at a low substrate temperature ($\sim$300$^\circ$C), conditions under which the Mn dopants randomly substitute for Ga throughout the ferromagnetic semiconductor layer \cite{Matsukara_JAP}. While we have not directly measured the Mn concentration in these samples, based upon the Mn:Ga beam equivalent flux ratio, past studies of Mn-doped GaAs, and the relatively low Curie temperature for ferromagnetic ordering ($T_C \sim 20$ K), we expect a Mn composition of a few percent. In samples referred to as ``type B,"  we directly grew a 60 nm thick Mn-doped GaSb buffer layer on the GaAs substrate at a significantly higher temperature (490$^\circ$C). Under these conditions, we expect to obtain a hybrid ferromagnet nanocluster (MnSb)/p-doped semiconductor (GaSb), allowing us to compare the behavior of Cd$_3$As$_2$ interfaced with a uniform ferromagnet to that of Cd$_3$As$_2$ interfaced with a highly nonuniform set of magnetic particles \cite{Jabolonska_doi:10.1063/1.3562171}. The final step in both type A and type B samples is to grow a Cd$_3$As$_2$ layer at a substrate temperature of $210^{\circ}$C. The thickness of these Cd$_3$As$_2$ layers ranged from 10 nm - 30 nm.   
 
 The thickness of each layer was determined by X-ray reflectivity (data not shown) and transmission electron microscopy (TEM). The surface roughness of the films was measured using atomic force microscopy (AFM). Magnetic properties of the ferromagnetic layer were investigated using superconducting quantum interference device (SQUID) magnetometry (Quantum Design MPMS3). 
To confirm that the Cd$_3$As$_2$ films indeed have DSM states after growth on Mn-doped GaSb, we used angle resolved photoemission spectroscopy (ARPES) measurements carried out via {\it in vacuo} transfer to a characterization chamber using a vacuum suitcase. Finally, we used PNR to obtain depth-resolved magnetization profiles in type A and B samples.

The growth of epitaxial Cd$_3$As$_2$ films on a Ga$_{1-x}$Mn$_x$Sb buffer layer resulted in a streaky RHEED pattern in both type A and type B samples, indicating smooth films on the length scale probed by RHEED (see Fig. 1 (a) for typical data for a type A film). Figure 1 (b) shows the out-of-plane high resolution X-ray diffraction (XRD) scans of a type A Cd$_3$As$_2$/Ga$_{1-x}$Mn$_x$Sb bilayer with peaks from (111) planes of the GaAs substrate and GaSb buffer layer which are marked with circles and triangles, respectively. Peaks arising from (112) planes of the Cd$_3$As$_2$ are indicated by a diamond shaped symbol. No additional peaks are present in the XRD spectrum, indicating single-phase GaSb and Cd$_3$As$_2$ films with a high degree of alignment in the out-of-plane direction. The insets show the rocking curves for the GaSb and Cd$_3$As$_2$ layers with full-width-half-maximum of about 0.04$^{\circ}$ and 0.11$^{\circ}$, respectively. The 5 {\textmu}m x 5 {\textmu}m AFM image of the Cd$_3$As$_2$ surface shows that the film is smooth with root mean square (RMS) roughness of 1.2 nm (Fig.1c). The  film surface is atomically stepped with step height of about 0.4 nm, which matches the interplanar spacing of the Cd$_3$As$_2$ (112) planes and is comparable to the values reported in the literature \cite{Schumann2017}. To gain further insight into the crystalline structure of our Ga$_{1-x}$Mn$_x$Sb and Cd$_3$As$_2$/Ga$_{1-x}$Mn$_x$Sb films, we carried out high-angle annular dark-field (HAADF) TEM measurements. These microscopy images show sharp interfaces between Cd$_3$As$_2$/GaSb and Cd$_3$As$_2$/Ga$_{1-x}$Mn$_x$Sb layers (Fig.1 (d) and 1 (e), respectively), with good crystalline quality of Cd$_3$As$_2$. The overview image shows sharp interfaces between each layer without intermixing (Inset of Fig.1 (e)). As expected, Type A heterostructures show relatively uniform Mn-distribution while Mn-rich nanoclusters suspected to be MnSb are observed in Type B samples, primarily near the Ga$_{1-x}$Mn$_x$Sb/Cd$_3$As$_2$ interface (see supplementary material \cite{Mitra_supp} and Fig. 7(c)). Our film characterization results show that we have successfully grown epitaxial Cd$_3$As$_2$ thin films on top of Ga$_{1-x}$Mn$_x$Sb with the appropriate crystal structure of the DSM phase.

To confirm the expected electronic structure and nontrivial topology of $\mathrm{Cd_3As_2}$, we performed ARPES measurements on type A Ga$_{1-x}$Mn$_x$Sb (Fig. 2 (a)) and type A $\mathrm{Cd_3As_2/Ga_{1-x}Mn_xSb}$ heterostructures (Fig. 2 (b),(c)). The scans were performed at room temperature along the $\overline{K}- \overline{\Gamma} - \overline{K}$ crystal momentum direction projected on the (111) surface for Ga$_{1-x}$Mn$_x$Sb and the (112) surface for $\mathrm{Cd_3As_2}$. The data in Fig. 2 were taken using the 21.1 eV $\mathrm{I\alpha}$ spectral line of a helium plasma lamp isolated by a monochromator. Additional data taken using the 7.2 eV and 8.4 eV (6th and 7th harmonics, respectively) of an ytterbium fiber vacuum ultra violet laser are shown in the Supplementary Materials \cite{Mitra_supp}. At room temperature, the Ga$_{1-x}$Mn$_x$Sb ARPES spectra show the chemical potential located above the valence band, confirming that the Mn-doping leads to p-type GaSb. We note that the p-doping prevents ARPES measurements from showing the full band gap. Figure 2 (b) shows the ARPES measurement of a type A $\mathrm{Cd_3As_2/Ga_{1-x}Mn_xSb}$ heterostructure at room temperature and Fig, 2 (c) shows its second derivative. This spectrum clearly show the characteristic linear dispersion of a DSM \cite{Liu2014,Borisenko2014,Yi2014,Yanez_PhysRevApplied.16.054031}.The ARPES data also show that the chemical potential of $\mathrm{Cd_3As_2}$ is located 0.2 eV above the charge neutral point. This is consistent with our previous work on $\mathrm{Cd_3As_2}$ grown on GaSb buffer layers and seems to indicate that the inclusion of Mn in GaSb does not affect the chemical potential or drastically change the band structure of the $\mathrm{Cd_3As_2}$ layer \cite{Yanez_PhysRevApplied.16.054031}. 

\begin{figure*}
\includegraphics[width=0.8\textwidth]{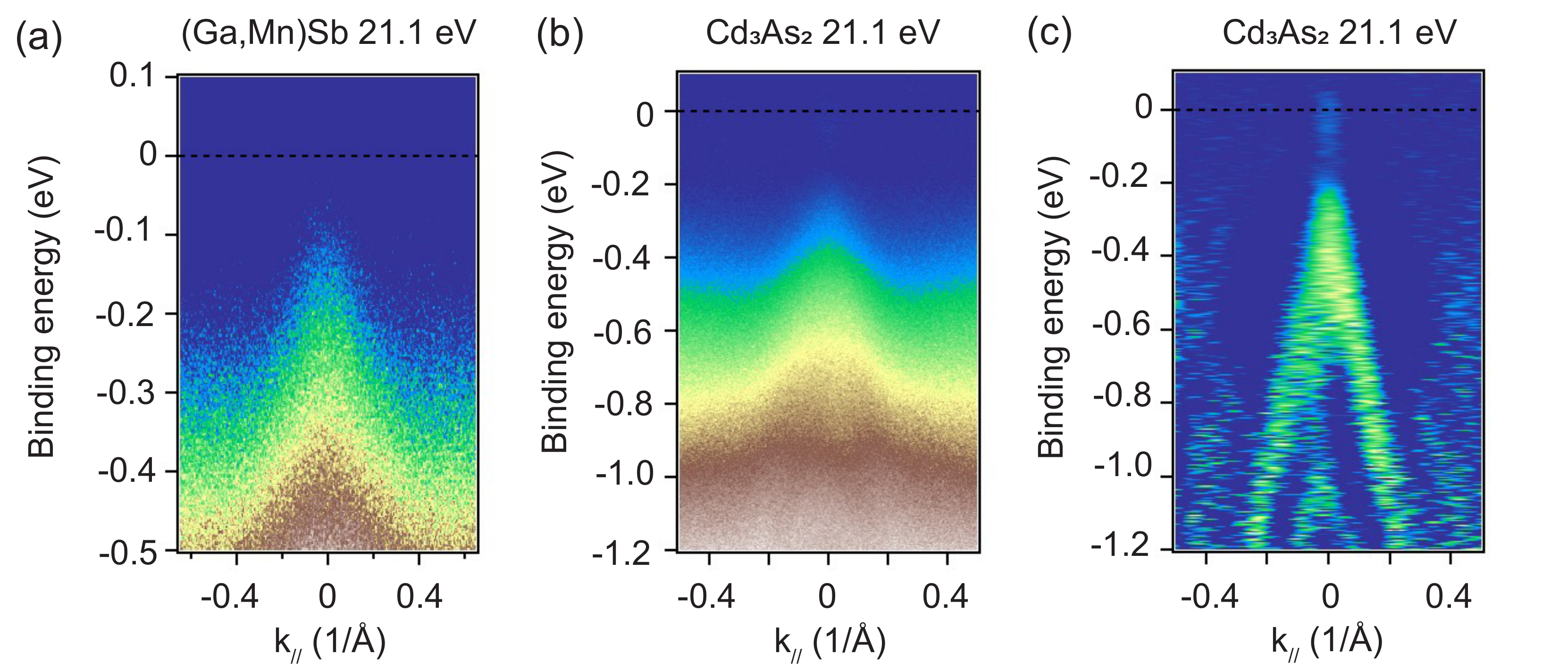}
\caption{\label{FIG.2} (a) ARPES data of Type A 10 nm thick Ga$_{1-x}$Mn$_x$Sb layer. The chemical potential in this p-doped sample is located 0.1 eV above the top of the valence band. (b) ARPES spectrum of a type A heterostructure consisting of 30 nm film of $\mathrm{Cd_3As_2}$ grown on 10 nm thick Ga$_{1-x}$Mn$_x$Sb. (c) Second derivative of the spectrum shown in panel (b). All scans were performed at 300 K using a photon excitation energy of 21.1 eV. The spectra are centered at the $\Gamma$ point and taken along the $\overline{K}- \overline{\Gamma} - \overline{K}$ direction projected on the (111) surface for Ga$_{1-x}$Mn$_x$Sb and the (112) surface for $\mathrm{Cd_3As_2}$.}
\end{figure*}


\section{\label{sec:level3}Magnetotransport measurements of the host materials}

To isolate signatures of emergent interfacial properties of the Cd$_3$As$_2$/(Ga,Mn)Sb heterostructures from those of the individual Cd$_3$As$_2$ and Ga$_{1-x}$Mn$_x$Sb films, we first probed the low temperature magnetotransport properties of Cd$_3$As$_2$/GaSb, type A Ga$_{1-x}$Mn$_x$Sb/GaSb, and type B Ga$_{1-x}$Mn$_x$Sb/GaSb. Magnetotransport measurements were performed in a Quantum Design Physical Property Measuring System (PPMS) at temperatures ranging from 2 K - 20 K. We used Ar{\text{+}} plasma etching to pattern the thin films into a standard Hall bar geometry with dimension 1 mm x 0.3 mm. 

The longitudinal sheet resistance (2D resistivity) $R_{xx}$ and Hall resistance $R_{xy}$ were measured using a bias current of 1 {\textmu}A with an out-of-plane magnetic field $B \leq 6$T. Following standard practice, all magnetic field dependent measurements of $R_{xx}$ ($R_{xy}$) were symmetrized (anti-symmetrized) with respect to the sign of the magnetic field. This is particularly important in the case of $R_{xy}$ since it removes cross-talk with contributions from $R_{xx}$.  

Figure 3 (a) shows $R_{xx}$ vs. $B$ applied normal to the sample plane in the temperature range 2 K to 10 K for a 15 nm thick Cd$_3$As$_2$/GaSb film. The Cd$_3$As$_2$ film exhibits a positive nonsaturating magnetoresistance (MR) which is likely a mixture of a classical parabolic MR and linear MR 
arising from charge density and/or mobility fluctuations \cite{Kisslinger2017,Parish2003,Kozlova2012}. At temperatures above 2 K, this Cd$_3$As$_2$ film does not show any obvious signatures of quantum corrections to diffusive transport, specifically a weak antilocalization (WAL) cusp in $R_{xx}$ vs. $B$ as the magnetic field is swept through $B = 0$. The variation of $R_{xy}$ vs. $B$ (Fig. 3 (b)) shows signs of multi-band conductivity in which the dominant carriers are n-type at low field, transitioning to a dominant p-type conduction for magnetic fields above approximately 3 T. This behavior has been attributed to samples wherein the chemical potential is near the bulk Dirac nodes in Cd$_3$As$_2$ \cite{Shoron_2021}.  We determined the mobility and carrier density \textit{n} of the dominant carriers using the linear low field region of the Hall data with a single band picture. In this region, \textit{n} is relatively insensitive to temperature, varying from 6.83 x 10$^{11}$ cm$^{-2}$ - 6.94 x 10$^{11}$ cm$^{-2}$ between 2 K and 10 K. For single crystals of Cd$_3$As$_2$, the corresponding three-dimensional carrier density is about an order of magnitude lower \cite{Neupane2014,Jeon2014}. We use the standard Drude model to determine a mobility {$\mu$} of about 10,500 cm$^{2}$ /(V.s). Note that the relatively high carrier density of the Cd$_3$As$_2$ is likely related to non-optimal growth conditions; more recent growth of Cd$_3$As$_2$ films in the same MBE chamber has
produced significantly lower carrier densities, resulting in the observation of Shubnikov de Haas oscilations and the integer quantum Hall at magnetic fields higher than 6 T \cite{Xiao_PhysRevB.106.L201101}. 

\begin{figure}
\includegraphics[width=0.8\textwidth]{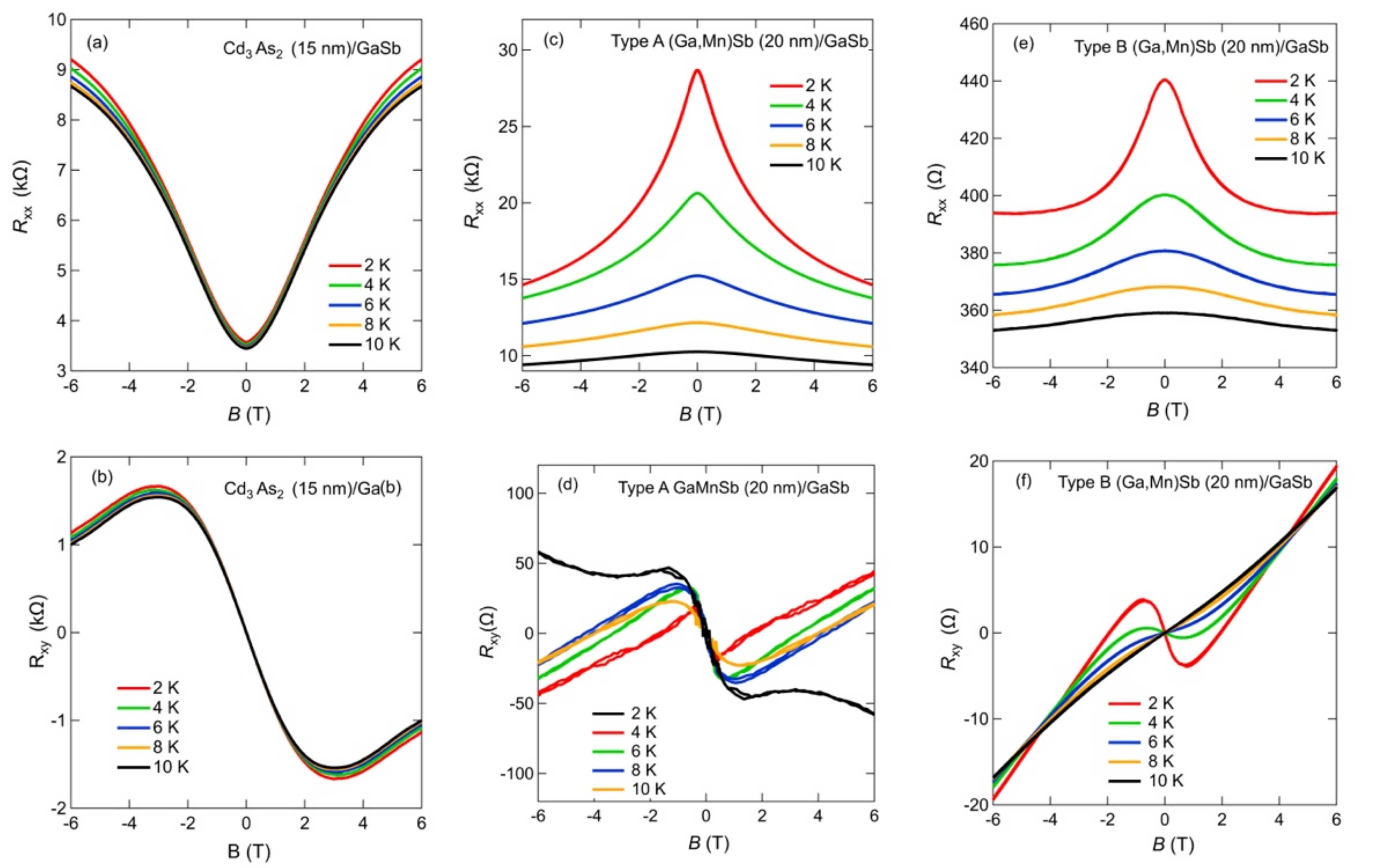}
\caption{\label{FIG.3} Top row: (a), (c), (e) Magnetic field dependence of $R_{xx}$ in a 15 nm thick Cd$_3$As$_2$ film, a 20 nm thick (Ga, Mn)Sb film, and a 20 nm thick type B (Ga, Mn)Sb film, respectively, in the temperature range 2 K $\leq T \leq 10$ K. Bottom row: (b), (d), (f) Magnetic field dependence of $R_{xy}$ for the same samples as in the upper row under identical conditions. All measurements are made in the standard Hall effect geometry ($\vec{B}$ perpendicular to both the current density and the sample plane). The longitudinal MR and Hall resistance measurements are symmetrized and antisymmetrized in magnetic field, respectively. No backgrounds have been subtracted}
\end{figure} 

We now describe the low temperature transport properties of a type A Ga$_{1-x}$Mn$_x$Sb (20 nm)/GaSb sample. Figure 3 (c) shows the magnetic field-dependence of $R_{xx}$ in the temperature range 2 K - 10 K, revealing a large negative MR attributed to the magnetic field suppression of spin disorder scattering, consistent with past observations in MBE-grown (Ga,Mn)Sb thin films \cite{Matsukara_JAP,MCCOMBE200390}. We note that, at low values of magnetic field, the resistivity of this type A Ga$_{1-x}$Mn$_x$Sb film is about an order of magnitude larger than that of the Cd$_3$As$_2$ film, so that most of the current flow in type A Cd$_3$As$_2$/Ga$_{1-x}$Mn$_x$Sb heterostructures will occur through the Cd$_3$As$_2$ layer. However, since the Cd$_3$As$_2$ and type A Ga$_{1-x}$Mn$_x$Sb films have MR of opposite sign, we anticipate a complex evolution of the current distribution in type A Cd$_3$As$_2$/Ga$_{1-x}$Mn$_x$Sb heterostructures at large applied magnetic fields. 

Figure 3 (d) shows the magnetic field-dependence of the Hall resistance, $R_{xy}$, in the type A Ga$_{1-x}$Mn$_x$Sb film over the temperature range 2 K - 10 K.  We caution that the Hall voltage, although properly antisymmetrized, is still significantly affected by cross-talk with the large longitudinal sheet resistance ($R_{xx} > 10 ~\mathrm{k}\Omega$), especially as the temperature is decreased below $\sim 6$ K.  We attribute the non-linear magnetic field dependence of $R_{xy}$ to an anomalous Hall effect arising from ferromagnetism in the hole-doped Ga$_{1-x}$Mn$_x$Sb layer. Indeed, the slope of $R_{xy} (B)$ at high magnetic fields shows that the film is p-type. The absence of any measurable hysteresis in these measurements with an out-of-plane magnetic field suggests an in-plane easy axis for the magnetization. We carried out magnetometry measurements on this film to gain further insight into the magnetism, with the caveat that reliable measurements of the magnetization in such thin, magnetically dilute films can be challenging to properly analyze because of the large diamagnetic background that has to be subtracted. The magnetic moment ($m$) versus $B$ data show very soft magnetic hysteresis loops for both in-plane and out-of-plane magnetic field orientation even at the lowest measurement temperature ($T=10$ K) \cite{Mitra_supp}. The coercive field (defined as value of $B$ where $m$ changes sign) is about 0.04 T and the magnetization of the sample fully saturates when $B \gtrsim 0.4$ T. Further, the magnetometry measurements show that magnetization has a relatively weak decrease with increasing temperature and a significant saturation moment remains even at room temperature ($T = 300$ K). These characteristics are similar to those reported in early studies of significantly thicker (200 nm) Ga$_{1-x}$Mn$_x$Sb films \cite{Matsukara_JAP}. We interpret our observations in the type A sample as indicative of a combination of hole-mediated ferromagnetism and nanosegregated Mn-rich regions (presumably MnSb) that are hard to detect in microscopy. Although we have not carried out a systematic set of measurements to determine the ferromagnetic transition temperatures ($T_C$) corresponding to these co-existing phases, the temperature dependence of the magnitude of the anomalous Hall effect suggests that the hole-mediated phase has a $T_C \sim 25$ K. The distinct change in the Hall effect for $T < 2$ K (Fig. 3 (d)) also suggests the onset of yet another low temperature phase.      

We next discuss the magnetotransport properties of a type B Ga$_{1-x}$Mn$_x$Sb(20 nm)/GaSb sample. The $R_{xx}$ vs. $B$ measurements of the type B Ga$_x$Mn$_{1-x}$Sb sample (Fig. 3 (e)) qualitatively resemble the behavior in type A Ga$_x$Mn$_{1-x}$Sb (Fig. 3 (d)), with strongly temperature-dependent negative longitudinal MR that becomes more pronounced as the temperature is lowered. However, there is an important difference: $R_{xx}$ in the type B sample is significantly smaller (almost two orders of magnitude) than in the type A sample. The $R_{xy}$ data for the type B sample (Fig. 3 (f)) indicates that the key difference is the much higher hole density, deduced from the slope of the linear behavior in $R_{xy}$ vs. $B$. This increased hole density is likely the result of Mn being incorporated more efficiently as an acceptor when Mn-doped GaSb is grown under more optimal conditions (higher substrate temperatures). SQUID magnetometry measurements of the type B Ga$_{1-x}$Mn$_x$Sb/GaSb sample show similar behavior to that of the type A sample, but with almost no hysteresis for both in-plane and out-of-plane field orientation (see Supplementary Materials \cite{Mitra_supp}). This indicates that the magnetization in the type B sample is dominated by a superparamagnetic contribution from ferromagnetic Mn-rich clusters seen in the TEM (Fig. S1 of the Supplementary Materials \cite{Mitra_supp}).    


\section{\label{sec:level4}Magnetotransport measurements of \texorpdfstring {C\MakeLowercase{d}$_3$A\MakeLowercase{s}$_2$}{Cd3As2}/\texorpdfstring {G\MakeLowercase{a}$_{1-x}$M\MakeLowercase{n}$_x$S\MakeLowercase{b}} {GaMnSb} heterostructures}

Having established the magnetotransport behavior of the constituent films, we now turn to the principle focus of this paper, namely the search for proximity-induced ferromagnetism in ferromagnetic semiconductor/DSM heterostructures. We begin with a discussion of magnetotransport measurements in a type A Cd$_3$As$_2$ (15 nm)/Ga$_{1-x}$Mn$_x$Sb (20 nm) sample.  

The sheet resistance ($R_{xx} \sim 2.7 ~ \mathrm{k}\Omega$  at 2 K), carrier density ($n \sim $ 1.8 x 10 $^{12}$ cm$^{-2}$ near $B = 0$), and mobility ($\mu \sim 6000$ cm$^{2}$/V-s) of the sample as deduced from measurements of $R_{xx}$ and $R_{xy}$ are similar to that of the individual Cd$_3$As$_2$ films, consistent with our expectation that the current in the type A heterostructure is dominated by the Cd$_3$As$_2$ channel. However, several key differences are obvious when we examine the variation of $R_{xx}$ and $R_{xy}$ with magnetic field and temperature (Figs. 4 (a) and (b), respectively). Figure 4 (a) shows that at higher temperatures ($T > 8$ K), the MR is positive, resembling that of the individual Cd$_3$As$_2$ film. As the temperature is lowered toward 2 K, there is a transition to a marked negative low field MR, taking on some of the characteristics of the type A Ga$_{1-x}$Mn$_x$Sb film. At all temperatures studied, the magnetic field dependence of the Hall effect is qualitatively similar to the non-linear dependence seen in the individual Cd$_3$As$_2$ film, but the non-linearity also has the marked temperature dependence of the the type A Ga$_{1-x}$Mn$_x$Sb film. We note that, in contrast, there is negligible temperature variation in the magnetic field dependence of $R_{xy}$ in the individual Cd$_3$As$_2$ film. These observations suggest that the electrical transport of charge carriers in the Cd$_3$As$_2$ channel is influenced by the ferromagnetism in the Ga$_{1-x}$Mn$_x$Sb channel, perhaps even picking up an anomalous Hall contribution. We caution though that this is not in itself an indication of proximity-induced ferromagnetic order in the Cd$_3$As$_2$ layer and could be the result of interfacial spin-dependent scattering of charge carriers in the Cd$_3$As$_2$. We also note that a detailed model and interpretation of transport in these heterostructures would have to account for the band realignment that occurs in what is essentially a p-n heterojunction (p-doped Ga$_{1-x}$Mn$_x$Sb and electron-dominated Cd$_3$As$_2$), an analysis of which is beyond the scope of this paper.          

\begin{figure}
\includegraphics[width=0.5\textwidth]{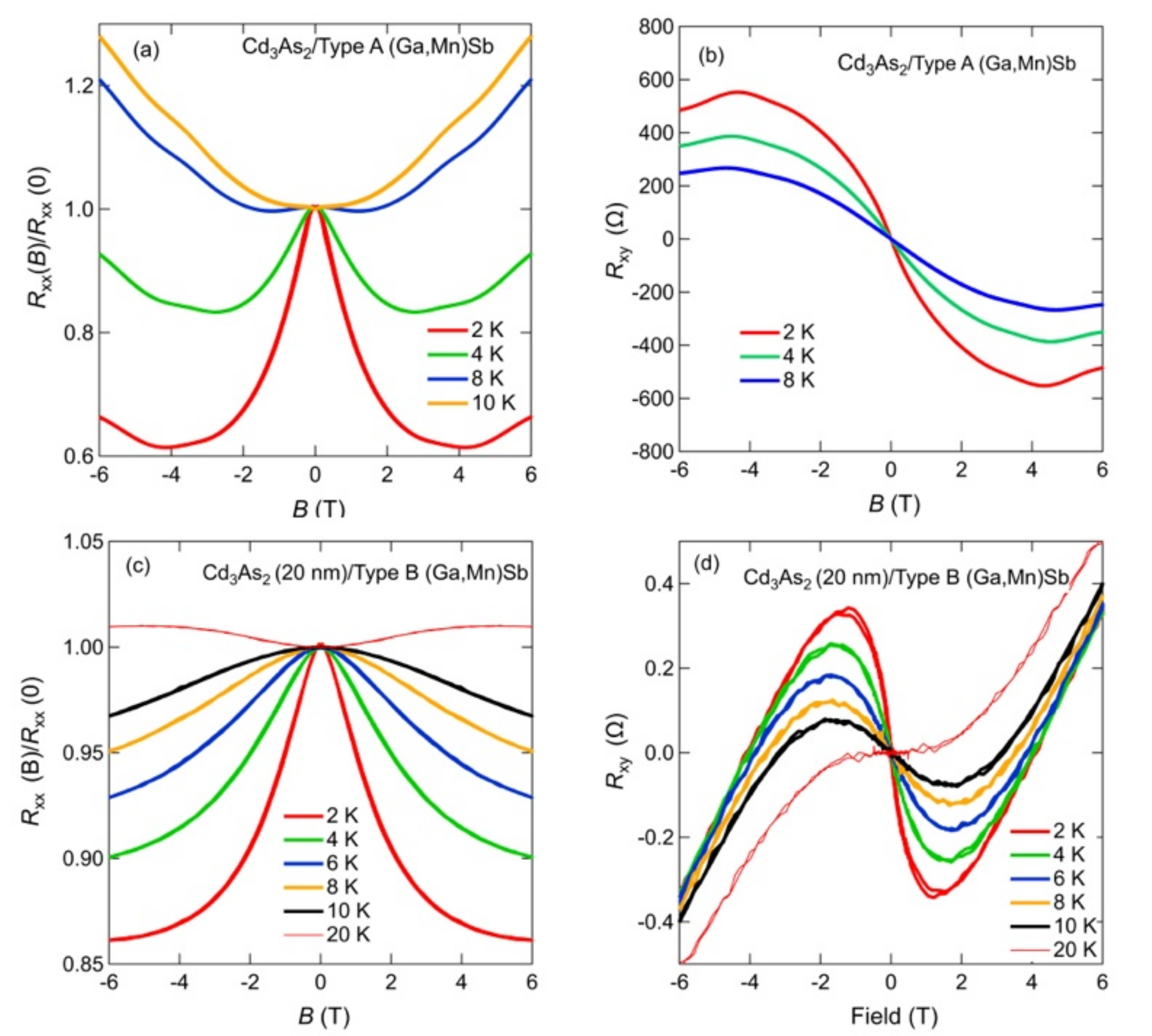}
\caption{\label{FIG.4} Magnetotransport properties of Cd$_3$As$_2$/Ga$_{1-x}$Mn$_x$Sb heterostructures. (a), (b) Magnetic field dependence of $R_{xx}$ and $R_{xy}$, respectively, in a type A Cd$_3$As$_2$/Ga$_{1-x}$Mn$_x$Sb Hall bar device in the temperature range 2 K $\leq T \leq 10$ K. (b),(c) Magnetic field dependence of $R_{xx}$ and $R_{xy}$, respectively, in a type B Cd$_3$As$_2$/Ga$_{1-x}$Mn$_x$Sb Hall bar device in the temperature range 2 K $\leq T \leq 20$ K. .}
\end{figure}


We now turn to a discussion of the magnetotransport in a type B Cd$_3$As$_2$/Ga$_x$Mn$_{1-x}$Sb heterostructure. As in the type A heterostructure, we observe a transition from a positive MR at higher temperature to a negative MR at low temperature (Fig. 4 (c)). The variation of $R_{xy}$ with magnetic field and temperature (Fig. 4 (d)) is qualitatively similar to that of the type B Ga$_x$Mn$_{1-x}$Sb film (Fig. 3 (f)). In these type B heterostructures, the low resistivity of the Ga$_x$Mn$_{1-x}$Sb presents a challenge for disentangling possible changes in the electrical transport in the Cd$_3$As$_2$ channel. Nonetheless, there is an interesting contrast between the behavior of the non-linear Hall data shown in Figs. 3 (f) and 4 (d).         

Comparing the electrical magnetotransport in type A and type B heterostructures, the most critical observation is that the type A samples show behavior that could be consistent with proximity-induced ferromagnetism, while in type B samples, interpretation is difficult in large part because of the comparable resistivities and thus parallel conduction in the DSM channel and the ferromagnetic semiconductor. The difference in magnetotransport behavior may be related to the uniformity of the magnetic interaction at the interface: the spin-polarized electrons in the ferromagnetic clusters near the interface cannot readily exert significant influence on charge carriers in the vicinal Cd$_3$As$_2$ layer via exchange or spin-dependent scattering. 



Although our magnetotransport measurements raise interesting questions about the interaction between the states of the DSM and the vicinal ferromagnet, it is difficult to distinguish between the different scenarios using transport measurements alone. To provide additional insight, we require a probe which is sensitive specifically to the interface. 
Consequently, we used PNR to probe the magnetization profile and searched for the presence of a magnetic proximity effect in Cd$_3$As$_2$ interfaced with Ga$_{1-x}$Mn$_x$Sb. 

\section{\label{sec:level5}PNR measurements of \texorpdfstring {C\MakeLowercase{d}$_3$A\MakeLowercase{s}$_2$}{Cd3As2}/\texorpdfstring {G\MakeLowercase{a}$_{1-x}$M\MakeLowercase{n}$_x$S\MakeLowercase{b}} {GaMnSb} heterostructures}

PNR data can be used to extract the chemical and in-plane net magnetization depth profile of the sample through fitting with a theoretical model. In PNR experiments, a spin-polarized neutron beam is incident on the sample, and the specular reflectivity is measured as a function of the momentum transfer vector along the film normal. The non-spin-flip scattering cross sections, in which the incident and scattered neutrons possess the same spin state, are sensitive to the depth profile of the net in-plane magnetization parallel to the applied field. On the other hand, the spin-flip scattering cross sections, in which the incident and scattered neutrons have opposite spins, are sensitive to the depth profile of the net in-plane magnetization perpendicular to the applied field.

The type A PNR measurement (Fig. 5) was performed using the Polarized Beam Reflectometer at the NIST Center for Neutron Research. Measurements were performed in full polarization mode, in which both the incident and scattered neutron spin is specified, at 6 K in an applied magnetic field of 3 T. Since this condition significantly exceeds the saturation fields shown in the magnetometry data \cite{Mitra_supp}, all net magnetization in the sample is expected to be aligned parallel to the field, such that the spin-flip scattering originating from any in-plane magnetization perpendicular to the field will be zero. Consequently, only the non-spin-flip cross sections were measured. The data were reduced using Reductus and analyzed using the Refl1D software package \cite{Maranville:po5131, KIRBY201244}. Models used to fit the data were constructed using a series of uniform slabs with different nuclear and magnetic scattering length densities to represent each layer in the structure. Slabs are separated by interfaces with Gaussian roughness, represented by an error function.

The best-fit solution (Fig. 5 (c)) consists of a net magnetization spread relatively uniformly throughout most of the Ga$_{1-x}$Mn$_x$Sb layer, with a total magnetic thickness of 12.5 nm $\pm$  1 nm. PNR estimates that 83\% $\pm$ 7\% of the Ga$_{1-x}$Mn$_x$Sb layer is magnetized, such that within a 95\% confidence interval essentially the entire layer may be considered ferromagnetic. In this case, then, the PNR results suggest that Mn-migration towards the interface with Cd$_3$As$_2$ is limited and the previously noted surfactant behaviors of Cd$_3$As$_2$ are significantly suppressed by the low growth temperature. The type A sample yields an interfacial magnetic SLD in the Cd$_3$As$_2$ of -0.10(7)$\times$10$^{-4}$ nm$^{-2}$, consistent with a complete lack of net magnetization within this layer. An upper limit of approximately 14 emu/cm$^3$ (1 emu/cm$^3$ = 1 kA/m) can be established for the proximity-induce magnetization at the interface with a 95\% confidence interval of the fitted PNR parameters. While such a small moment is within the uncertainty of this PNR measurement, it may be sufficient to break time reversal symmetry in the Cd$_3$As$_2$ at the interface.

\begin{figure}
\includegraphics[width=0.5\textwidth]{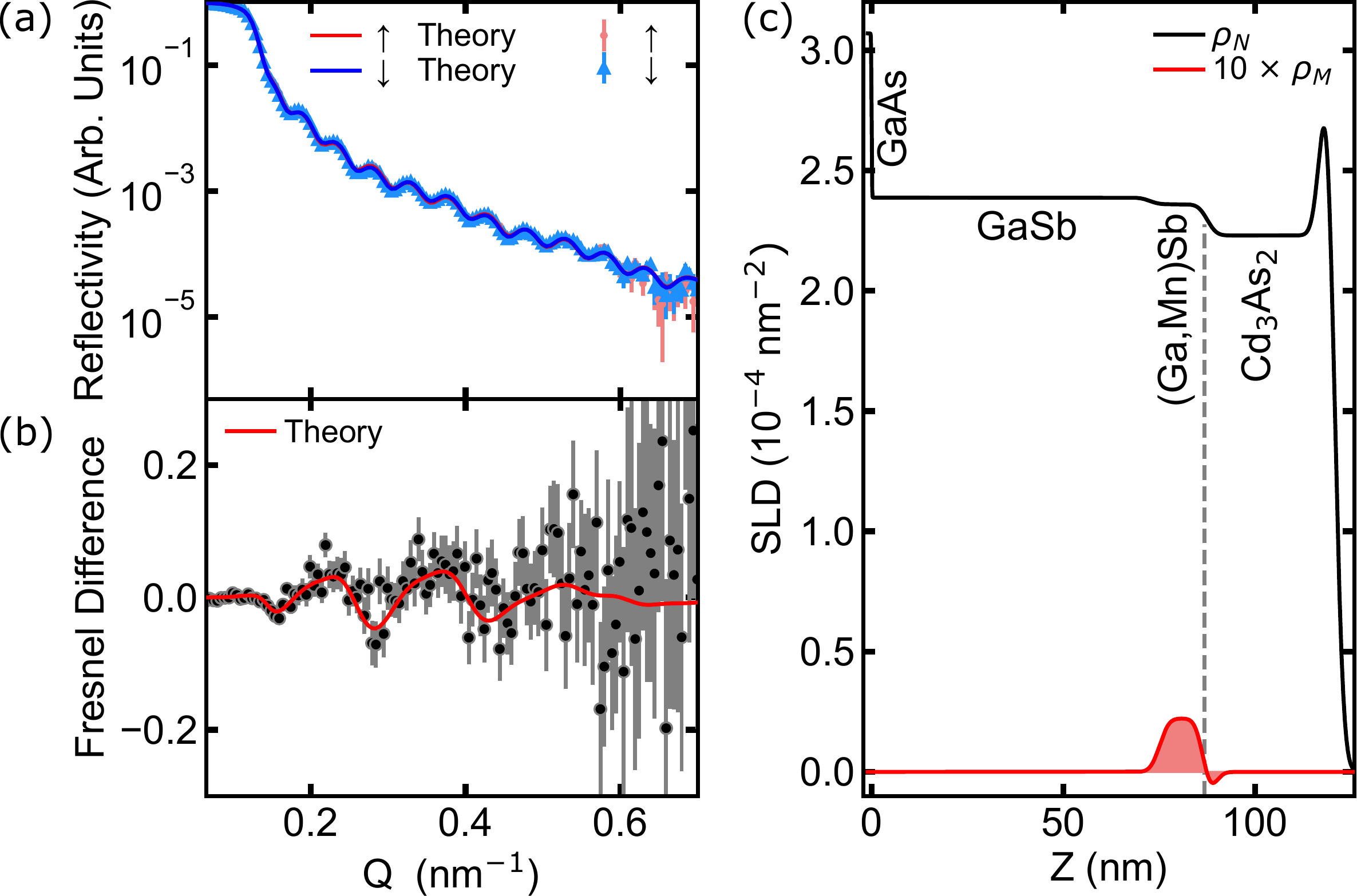}
\caption{\label{FIG.6} (a) Spin-dependent polarized neutron reflectivity measurements of a type A sample alongside theoretical fits. (b) Difference between the spin-down and spin-up neutron reflectivities shown in (a) normalized by the theoretical reflectivity of the GaAs substrate. (c) Nuclear and magnetic scattering length density profiles used to generate the fits shown. Measurements were performed at 6 K in an applied magnetic field of 3 T. Error bars represent $\pm$1 standard deviation.}
\end{figure}

PNR measurements on a type B sample were performed using the MAGREF instrument at the Spallation Neutron Source at a temperature of 5 K in an in-plane applied magnetic field of 1.15 T. Once again this exceeds the field necessary to saturate the sample, so that no net perpendicular magnetization is expected. We therefore measured the spin-dependent neutron reflectivities as a function of the perpendicular momentum transfer vector \textbf{Q} without polarization analysis of the scattered beam. We collected the spin-up and spin-down reflectivities under the assumption that these cross sections contained only non-spin-flip scattering. The data were reduced using a custom Python code, and again fit using Refl1D. Since MAGREF is a time-of-flight neutron reflectometer, sections of the PNR data were taken at a series of scattering angles, each of which captures data across a range of Q-values. Each of these angular regions was allowed some variation in relative intensity and angle of incidence during the fit to compensate for known imperfections in automated data stitching programs. Data were then rebinned after analysis to a continuous $\Delta$Q/Q of 0.025. 

Interestingly, fitting the PNR from the type B sample was significantly more complex than that of type A, most likely due to the Mn diffusion and agglomeration expected at the higher growth temperature. To improve modeling constraints and ameliorate the relatively weak contrast between the nuclear scattering length densities of bulk GaSb, Ga$_{1-x}$Mn$_x$Sb, and Cd$_3$As$_2$, the sample used for this measurement was subsequently examined by STEM imaging which allowed the relative thicknesses of all three layers to be determined directly. This thickness ratio was implemented as a constraint in the modeling to enable robust conclusions to be drawn about the position of the magnetization within the stack. We considered the following possible magnetic models:

\begin{enumerate}

\item All magnetization is concentrated at the Ga$_{1-x}$Mn$_x$Sb/Cd$_3$As$_2$ interface.
\item All magnetization is uniformly distributed throughout the Ga$_{1-x}$Mn$_x$Sb layer.
\item All magnetization is uniformly distributed throughout the Cd$_3$As$_2$ layer.
\item The magnetization is distributed throughout both the Ga$_{1-x}$Mn$_x$Sb and Cd$_3$As$_2$.
\item The magnetization is confined to a Mn-rich layer on the top Cd$_3$As$_2$ surface.
\item The magnetization is present in both the Ga$_{1-x}$Mn$_x$Sb and a Mn-rich surface layer on top of the Cd$_3$As$_2$.
\end{enumerate}

Of these models, only $\# 1$ is able to properly describe the data, as shown in Fig. 6. Figure 6 (a) shows the fitted spin-dependent reflectivities while Fig. 6 (b) plots the difference between the two curves normalized by the Fresnel reflectivity alongside a theoretical curve, which we refer to as the ``Fresnel difference''. Here, the Frensel reflectivity is defined as the reflectivity expected purely from the bare GaAs substrate used in this experiment. As shown in the plot of the corresponding depth-dependent structural and magnetic scattering length densities (SLDs) that generated the fit (Fig. 6 (c)),  the PNR revealed high-quality films with sharp interfaces and SLDs near theoretically expected bulk values.

Fits from the alternative models ($\# 2- \# 6$), which fail to provide a reasonable fit to the data, are shown in the Supplementary Information. Note that models 1, 3, and 5 were designed to allow a magnetic proximity effect in which a portion of the net magnetization extends into the Cd$_3$As$_2$. In fact, for the best-fit model $\# 1$, we find that the magnetization at the interface extends into both the (Ga,Mn)As and Cd$_3$As$_2$ layers (Fig. 6 (c)). Since the depth-dependent SLD profiles obtained from PNR represent an average across the sample plane, this result is consistent with either the formation of ferromagnetic Mn-rich inclusions directly at the interface observed in the associated STEM imaging (see also Supplementary Information Fig. S1), or a magnetic proximity effect in the Cd$_3$As$_2$. Separating proximity induced magnetism from interfacial Mn-rich phases based on the PNR alone is extremely challenging, but the transport results strongly favor the former. Thus, we conclude that the larger magnetization tightly constrained to the interface is a result of ferromagnetic MnSb inclusions.

\begin{figure}
\includegraphics[width=0.5\textwidth]{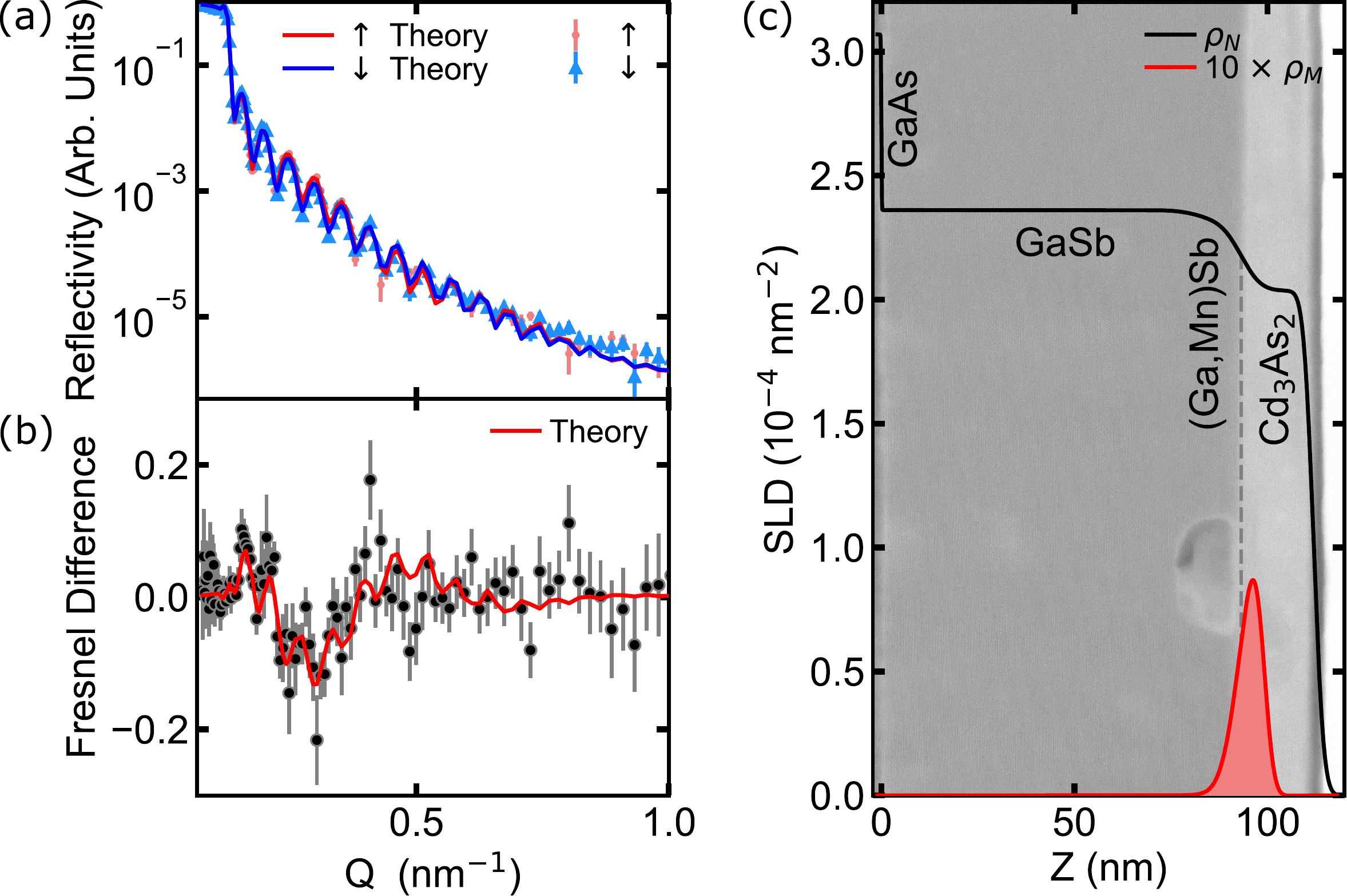}
\caption{\label{FIG.7} (a) Spin-dependent polarized neutron reflectivity measurements of a type B sample alongside theoretical fits. (b) Difference between the spin-down and spin-up neutron reflectivities shown in (a) normalized by the theoretical reflectivity of the GaAs substrate. (c) Nuclear and magnetic scattering length density profiles used to generate the fits shown. The data are superimposed on a TEM image of this type B sample. Measurements were performed at 5 K in an applied magnetic field of 1.15 T. Error bars represent $\pm$1 standard deviation.}
\end{figure}

For a direct comparison, we also optimized a model depth profile for the type A sample similar to Model $\# 1$ used for the type B sample, which yielded nonphysical depth profiles and resulted in a much worse fit to the data. Thus PNR demonstrates stark differences in the Mn distribution and resulting magnetic state of type A and type B samples.

\section{\label{sec:level6}Conclusions}

In summary, we have reported a comprehensive study of the epitaxial synthesis, crystal structure, surface band structure, electrical magneto-transport, and magnetic behavior of an archetypal DSM ($\mathrm{Cd_3 As_2}$)/ferromagnetic semiconductor ($\mathrm{Ga_{1-x}Mn_x Sb}$) heterostructure. Two growth regimes were investigated for the ferromagnetic semiconductor, with a type A sample that has a more uniform incorporation of Mn and a type B sample that has possible clustering of Mn into ferromagnetic phases (such as MnSb). At first glance, the variation with magnetic field and temperature of the longitudinal MR and Hall effect in the type A samples suggests the possibility of a proximity-induced ferromagnetic effect in this bilayer. However, while PNR indicates that ferromagnetic order in the type A sample, it is only seen in the Ga$_{1-x}$Mn$_x$Sb layer and any induced magnetization in the Cd$_3$As$_2$ layer is expected to be smaller than 14 emu/cm$^{3}$ (95\% confidence limit). We conclude therefore that the magnetotransport signatures in type A samples originate in spin-dependent scattering of charge carriers in the Cd$_3$As$_2$ channel and cannot be attributed to a magnetic proximity effect, at least at temperatures above about 4 K. This does not rule out the possibility of a proximity effect being responsible for the observed hysteretic MR and Hall effect at lower temperatures $T \leq 2$ K. In a type B sample, the totality of PNR, STEM, and transport evidence reveals the presence of ferromagnetic Mn-rich inclusions (most likely MnSb) which segregate towards the interface region. While this also puts Cd$_3$As$_2$ in direct contact with a ferromagnetic phase, the uniformity of the ferromagnet is greatly reduced and there is no evidence of a magnetic proximity effect or broken TRS in the magnetotransport. We consequently conclude that, while small local regions of the Cd$_3$As$_2$ may experience the effects of spin-polarization at the interface, most of the bulk conduction channels are unaffected by the inhomogeneous magnetism of type B (Ga$_x$Mn$_{1-x}$)Sb. Overall, our results support the importance of maintaining a uniformly magnetized, high-quality ferromagnetic interface in order to observe the effects of magnetic proximity in Dirac semimetal heterostructures. 

\begin{acknowledgments}
This work was supported by the Institute for Quantum Matter under DOE EFRC grant DE-SC0019331 (AM, RX, JC, TM, NS). The Penn State Two-Dimensional Crystal Consortium Materials Innovation Platform (2DCC-MIP) under NSF Grant No. DMR-2039351 provided support for ARPES measurements (YO, NS). WY acknowledges support from the Penn State Materials Research Science and Engineering Center for Nanoscale Science under NSF Grant No. DMR-2011839. Certain commercial products or company names are identified here to describe our study adequately. Such identification is not intended to imply recommendation or endorsement by the National Institute of Standards and Technology, nor is it intended to imply that the products or names identified are necessarily the best available for the purpose. 
\end{acknowledgments}

\nocite{*}

\begin{thebibliography}{62}%
\makeatletter
\providecommand \@ifxundefined [1]{%
 \@ifx{#1\undefined}
}%
\providecommand \@ifnum [1]{%
 \ifnum #1\expandafter \@firstoftwo
 \else \expandafter \@secondoftwo
 \fi
}%
\providecommand \@ifx [1]{%
 \ifx #1\expandafter \@firstoftwo
 \else \expandafter \@secondoftwo
 \fi
}%
\providecommand \natexlab [1]{#1}%
\providecommand \enquote  [1]{``#1''}%
\providecommand \bibnamefont  [1]{#1}%
\providecommand \bibfnamefont [1]{#1}%
\providecommand \citenamefont [1]{#1}%
\providecommand \href@noop [0]{\@secondoftwo}%
\providecommand \href [0]{\begingroup \@sanitize@url \@href}%
\providecommand \@href[1]{\@@startlink{#1}\@@href}%
\providecommand \@@href[1]{\endgroup#1\@@endlink}%
\providecommand \@sanitize@url [0]{\catcode `\\12\catcode `\$12\catcode
  `\&12\catcode `\#12\catcode `\^12\catcode `\_12\catcode `\%12\relax}%
\providecommand \@@startlink[1]{}%
\providecommand \@@endlink[0]{}%
\providecommand \url  [0]{\begingroup\@sanitize@url \@url }%
\providecommand \@url [1]{\endgroup\@href {#1}{\urlprefix }}%
\providecommand \urlprefix  [0]{URL }%
\providecommand \Eprint [0]{\href }%
\providecommand \doibase [0]{https://doi.org/}%
\providecommand \selectlanguage [0]{\@gobble}%
\providecommand \bibinfo  [0]{\@secondoftwo}%
\providecommand \bibfield  [0]{\@secondoftwo}%
\providecommand \translation [1]{[#1]}%
\providecommand \BibitemOpen [0]{}%
\providecommand \bibitemStop [0]{}%
\providecommand \bibitemNoStop [0]{.\EOS\space}%
\providecommand \EOS [0]{\spacefactor3000\relax}%
\providecommand \BibitemShut  [1]{\csname bibitem#1\endcsname}%
\let\auto@bib@innerbib\@empty
\bibitem [{\citenamefont {Hauser}(1969)}]{Hauser_PR_1969}%
  \BibitemOpen
  \bibfield  {author} {\bibinfo {author} {\bibfnamefont {J.~J.}\ \bibnamefont
  {Hauser}},\ }\href {https://doi.org/10.1103/PhysRev.187.580} {\bibfield
  {journal} {\bibinfo  {journal} {Phys. Rev.}\ }\textbf {\bibinfo {volume}
  {187}},\ \bibinfo {pages} {580} (\bibinfo {year} {1969})}\BibitemShut
  {NoStop}%
\bibitem [{\citenamefont {Huang}\ \emph {et~al.}(2012)\citenamefont {Huang},
  \citenamefont {Fan}, \citenamefont {Qu}, \citenamefont {Chen}, \citenamefont
  {Wang}, \citenamefont {Wu}, \citenamefont {Chen}, \citenamefont {Xiao},\ and\
  \citenamefont {Chien}}]{Huang_PhysRevLett.109.107204}%
  \BibitemOpen
  \bibfield  {author} {\bibinfo {author} {\bibfnamefont {S.~Y.}\ \bibnamefont
  {Huang}}, \bibinfo {author} {\bibfnamefont {X.}~\bibnamefont {Fan}}, \bibinfo
  {author} {\bibfnamefont {D.}~\bibnamefont {Qu}}, \bibinfo {author}
  {\bibfnamefont {Y.~P.}\ \bibnamefont {Chen}}, \bibinfo {author}
  {\bibfnamefont {W.~G.}\ \bibnamefont {Wang}}, \bibinfo {author}
  {\bibfnamefont {J.}~\bibnamefont {Wu}}, \bibinfo {author} {\bibfnamefont
  {T.~Y.}\ \bibnamefont {Chen}}, \bibinfo {author} {\bibfnamefont {J.~Q.}\
  \bibnamefont {Xiao}},\ and\ \bibinfo {author} {\bibfnamefont {C.~L.}\
  \bibnamefont {Chien}},\ }\href
  {https://doi.org/10.1103/PhysRevLett.109.107204} {\bibfield  {journal}
  {\bibinfo  {journal} {Phys. Rev. Lett.}\ }\textbf {\bibinfo {volume} {109}},\
  \bibinfo {pages} {107204} (\bibinfo {year} {2012})}\BibitemShut {NoStop}%
\bibitem [{\citenamefont {Vobornik}\ \emph {et~al.}(2011)\citenamefont
  {Vobornik}, \citenamefont {Manju}, \citenamefont {Fujii}, \citenamefont
  {Borgatti}, \citenamefont {Torelli}, \citenamefont {Krizmancic},
  \citenamefont {Hor}, \citenamefont {Cava},\ and\ \citenamefont
  {Panaccione}}]{Vobornik_WOS:000295667000009}%
  \BibitemOpen
  \bibfield  {author} {\bibinfo {author} {\bibfnamefont {I.}~\bibnamefont
  {Vobornik}}, \bibinfo {author} {\bibfnamefont {U.}~\bibnamefont {Manju}},
  \bibinfo {author} {\bibfnamefont {J.}~\bibnamefont {Fujii}}, \bibinfo
  {author} {\bibfnamefont {F.}~\bibnamefont {Borgatti}}, \bibinfo {author}
  {\bibfnamefont {P.}~\bibnamefont {Torelli}}, \bibinfo {author} {\bibfnamefont
  {D.}~\bibnamefont {Krizmancic}}, \bibinfo {author} {\bibfnamefont {Y.~S.}\
  \bibnamefont {Hor}}, \bibinfo {author} {\bibfnamefont {R.~J.}\ \bibnamefont
  {Cava}},\ and\ \bibinfo {author} {\bibfnamefont {G.}~\bibnamefont
  {Panaccione}},\ }\href {https://doi.org/10.1021/nl201275q} {\bibfield
  {journal} {\bibinfo  {journal} {Nano Lett.}\ }\textbf {\bibinfo {volume}
  {11}},\ \bibinfo {pages} {4079} (\bibinfo {year} {2011})}\BibitemShut
  {NoStop}%
\bibitem [{\citenamefont {Lee}\ \emph {et~al.}(2018)\citenamefont {Lee},
  \citenamefont {Richardella}, \citenamefont {Fraleigh}, \citenamefont {Liu},
  \citenamefont {Zhao},\ and\ \citenamefont {Samarth}}]{JSLee_npjQM}%
  \BibitemOpen
  \bibfield  {author} {\bibinfo {author} {\bibfnamefont {J.~S.}\ \bibnamefont
  {Lee}}, \bibinfo {author} {\bibfnamefont {A.}~\bibnamefont {Richardella}},
  \bibinfo {author} {\bibfnamefont {R.~D.}\ \bibnamefont {Fraleigh}}, \bibinfo
  {author} {\bibfnamefont {C.-x.}\ \bibnamefont {Liu}}, \bibinfo {author}
  {\bibfnamefont {W.}~\bibnamefont {Zhao}},\ and\ \bibinfo {author}
  {\bibfnamefont {N.}~\bibnamefont {Samarth}},\ }\href
  {https://www.nature.com/articles/s41535-018-0123-2#citeas} {\bibfield
  {journal} {\bibinfo  {journal} {npj Quantum Mater.}\ }\textbf {\bibinfo
  {volume} {3}},\ \bibinfo {pages} {51} (\bibinfo {year} {2018})}\BibitemShut
  {NoStop}%
\bibitem [{\citenamefont {Tang}\ \emph {et~al.}(2017)\citenamefont {Tang},
  \citenamefont {Chang}, \citenamefont {Zhao}, \citenamefont {Liu},
  \citenamefont {Jiang}, \citenamefont {Liu}, \citenamefont {McCartney},
  \citenamefont {Smith}, \citenamefont {Chen}, \citenamefont {Moodera},\ and\
  \citenamefont {Shi}}]{Tang_WOS:000406370700063}%
  \BibitemOpen
  \bibfield  {author} {\bibinfo {author} {\bibfnamefont {C.}~\bibnamefont
  {Tang}}, \bibinfo {author} {\bibfnamefont {C.-Z.}\ \bibnamefont {Chang}},
  \bibinfo {author} {\bibfnamefont {G.}~\bibnamefont {Zhao}}, \bibinfo {author}
  {\bibfnamefont {Y.}~\bibnamefont {Liu}}, \bibinfo {author} {\bibfnamefont
  {Z.}~\bibnamefont {Jiang}}, \bibinfo {author} {\bibfnamefont {C.-X.}\
  \bibnamefont {Liu}}, \bibinfo {author} {\bibfnamefont {M.~R.}\ \bibnamefont
  {McCartney}}, \bibinfo {author} {\bibfnamefont {D.~J.}\ \bibnamefont
  {Smith}}, \bibinfo {author} {\bibfnamefont {T.}~\bibnamefont {Chen}},
  \bibinfo {author} {\bibfnamefont {J.~S.}\ \bibnamefont {Moodera}},\ and\
  \bibinfo {author} {\bibfnamefont {J.}~\bibnamefont {Shi}},\ }\href
  {https://doi.org/10.1126/sciadv.1700307} {\bibfield  {journal} {\bibinfo
  {journal} {Sci. Adv.}\ }\textbf {\bibinfo {volume} {3}},\ \bibinfo {pages}
  {e1700307} (\bibinfo {year} {2017})}\BibitemShut {NoStop}%
\bibitem [{\citenamefont {Katmis}\ \emph {et~al.}(2016)\citenamefont {Katmis},
  \citenamefont {Lauter}, \citenamefont {Nogueira}, \citenamefont {Assaf},
  \citenamefont {Jamer}, \citenamefont {Wei}, \citenamefont {Satpati},
  \citenamefont {Freeland}, \citenamefont {Eremin}, \citenamefont {Heiman},
  \citenamefont {Jarillo-Herrero},\ and\ \citenamefont
  {Moodera}}]{Katmis_WOS:000376443100036}%
  \BibitemOpen
  \bibfield  {author} {\bibinfo {author} {\bibfnamefont {F.}~\bibnamefont
  {Katmis}}, \bibinfo {author} {\bibfnamefont {V.}~\bibnamefont {Lauter}},
  \bibinfo {author} {\bibfnamefont {F.~S.}\ \bibnamefont {Nogueira}}, \bibinfo
  {author} {\bibfnamefont {B.~A.}\ \bibnamefont {Assaf}}, \bibinfo {author}
  {\bibfnamefont {M.~E.}\ \bibnamefont {Jamer}}, \bibinfo {author}
  {\bibfnamefont {P.}~\bibnamefont {Wei}}, \bibinfo {author} {\bibfnamefont
  {B.}~\bibnamefont {Satpati}}, \bibinfo {author} {\bibfnamefont {J.~W.}\
  \bibnamefont {Freeland}}, \bibinfo {author} {\bibfnamefont {I.}~\bibnamefont
  {Eremin}}, \bibinfo {author} {\bibfnamefont {D.}~\bibnamefont {Heiman}},
  \bibinfo {author} {\bibfnamefont {P.}~\bibnamefont {Jarillo-Herrero}},\ and\
  \bibinfo {author} {\bibfnamefont {J.~S.}\ \bibnamefont {Moodera}},\ }\href
  {https://doi.org/10.1038/nature17635} {\bibfield  {journal} {\bibinfo
  {journal} {Nature}\ }\textbf {\bibinfo {volume} {533}},\ \bibinfo {pages}
  {513+} (\bibinfo {year} {2016})}\BibitemShut {NoStop}%
\bibitem [{\citenamefont {Riddiford}\ \emph
  {et~al.}(2022{\natexlab{a}})\citenamefont {Riddiford}, \citenamefont
  {Grutter}, \citenamefont {Pillsbury}, \citenamefont {Stanley}, \citenamefont
  {Reifsnyder~Hickey}, \citenamefont {Li}, \citenamefont {Alem}, \citenamefont
  {Samarth},\ and\ \citenamefont {Suzuki}}]{Riddiford_PhysRevLett.128.126802}%
  \BibitemOpen
  \bibfield  {author} {\bibinfo {author} {\bibfnamefont {L.~J.}\ \bibnamefont
  {Riddiford}}, \bibinfo {author} {\bibfnamefont {A.~J.}\ \bibnamefont
  {Grutter}}, \bibinfo {author} {\bibfnamefont {T.}~\bibnamefont {Pillsbury}},
  \bibinfo {author} {\bibfnamefont {M.}~\bibnamefont {Stanley}}, \bibinfo
  {author} {\bibfnamefont {D.}~\bibnamefont {Reifsnyder~Hickey}}, \bibinfo
  {author} {\bibfnamefont {P.}~\bibnamefont {Li}}, \bibinfo {author}
  {\bibfnamefont {N.}~\bibnamefont {Alem}}, \bibinfo {author} {\bibfnamefont
  {N.}~\bibnamefont {Samarth}},\ and\ \bibinfo {author} {\bibfnamefont
  {Y.}~\bibnamefont {Suzuki}},\ }\href
  {https://doi.org/10.1103/PhysRevLett.128.126802} {\bibfield  {journal}
  {\bibinfo  {journal} {Phys. Rev. Lett.}\ }\textbf {\bibinfo {volume} {128}},\
  \bibinfo {pages} {126802} (\bibinfo {year} {2022}{\natexlab{a}})}\BibitemShut
  {NoStop}%
\bibitem [{\citenamefont {Crassee}\ \emph {et~al.}(2018)\citenamefont
  {Crassee}, \citenamefont {Sankar}, \citenamefont {Lee}, \citenamefont
  {Akrap},\ and\ \citenamefont {Orlita}}]{Crassee}%
  \BibitemOpen
  \bibfield  {author} {\bibinfo {author} {\bibfnamefont {I.}~\bibnamefont
  {Crassee}}, \bibinfo {author} {\bibfnamefont {R.}~\bibnamefont {Sankar}},
  \bibinfo {author} {\bibfnamefont {W.-L.}\ \bibnamefont {Lee}}, \bibinfo
  {author} {\bibfnamefont {A.}~\bibnamefont {Akrap}},\ and\ \bibinfo {author}
  {\bibfnamefont {M.}~\bibnamefont {Orlita}},\ }\href
  {https://doi.org/10.1103/PhysRevMaterials.2.120302} {\bibfield  {journal}
  {\bibinfo  {journal} {Phys. Rev. Materials}\ }\textbf {\bibinfo {volume}
  {2}},\ \bibinfo {pages} {120302} (\bibinfo {year} {2018})}\BibitemShut
  {NoStop}%
\bibitem [{\citenamefont {Cano}\ \emph {et~al.}(2017)\citenamefont {Cano},
  \citenamefont {Bradlyn}, \citenamefont {Wang}, \citenamefont {Hirschberger},
  \citenamefont {Ong},\ and\ \citenamefont
  {Bernevig}}]{Cano_PhysRevB.95.161306}%
  \BibitemOpen
  \bibfield  {author} {\bibinfo {author} {\bibfnamefont {J.}~\bibnamefont
  {Cano}}, \bibinfo {author} {\bibfnamefont {B.}~\bibnamefont {Bradlyn}},
  \bibinfo {author} {\bibfnamefont {Z.}~\bibnamefont {Wang}}, \bibinfo {author}
  {\bibfnamefont {M.}~\bibnamefont {Hirschberger}}, \bibinfo {author}
  {\bibfnamefont {N.~P.}\ \bibnamefont {Ong}},\ and\ \bibinfo {author}
  {\bibfnamefont {B.~A.}\ \bibnamefont {Bernevig}},\ }\href
  {https://doi.org/10.1103/PhysRevB.95.161306} {\bibfield  {journal} {\bibinfo
  {journal} {Phys. Rev. B}\ }\textbf {\bibinfo {volume} {95}},\ \bibinfo
  {pages} {161306} (\bibinfo {year} {2017})}\BibitemShut {NoStop}%
\bibitem [{\citenamefont {Baidya}\ and\ \citenamefont
  {Vanderbilt}(2020)}]{Baidya_PhysRevB.102.165115}%
  \BibitemOpen
  \bibfield  {author} {\bibinfo {author} {\bibfnamefont {S.}~\bibnamefont
  {Baidya}}\ and\ \bibinfo {author} {\bibfnamefont {D.}~\bibnamefont
  {Vanderbilt}},\ }\href {https://doi.org/10.1103/PhysRevB.102.165115}
  {\bibfield  {journal} {\bibinfo  {journal} {Phys. Rev. B}\ }\textbf {\bibinfo
  {volume} {102}},\ \bibinfo {pages} {165115} (\bibinfo {year}
  {2020})}\BibitemShut {NoStop}%
\bibitem [{\citenamefont {Xiao}\ \emph
  {et~al.}(2022{\natexlab{a}})\citenamefont {Xiao}, \citenamefont {Held},
  \citenamefont {Rable}, \citenamefont {Ghosh}, \citenamefont {Wang},
  \citenamefont {Mkhoyan},\ and\ \citenamefont
  {Samarth}}]{Xiao_PhysRevMaterials.6.024203}%
  \BibitemOpen
  \bibfield  {author} {\bibinfo {author} {\bibfnamefont {R.}~\bibnamefont
  {Xiao}}, \bibinfo {author} {\bibfnamefont {J.~T.}\ \bibnamefont {Held}},
  \bibinfo {author} {\bibfnamefont {J.}~\bibnamefont {Rable}}, \bibinfo
  {author} {\bibfnamefont {S.}~\bibnamefont {Ghosh}}, \bibinfo {author}
  {\bibfnamefont {K.}~\bibnamefont {Wang}}, \bibinfo {author} {\bibfnamefont
  {K.~A.}\ \bibnamefont {Mkhoyan}},\ and\ \bibinfo {author} {\bibfnamefont
  {N.}~\bibnamefont {Samarth}},\ }\href
  {https://doi.org/10.1103/PhysRevMaterials.6.024203} {\bibfield  {journal}
  {\bibinfo  {journal} {Phys. Rev. Materials}\ }\textbf {\bibinfo {volume}
  {6}},\ \bibinfo {pages} {024203} (\bibinfo {year}
  {2022}{\natexlab{a}})}\BibitemShut {NoStop}%
\bibitem [{\citenamefont {Liu}\ \emph {et~al.}(2014)\citenamefont {Liu},
  \citenamefont {Jiang}, \citenamefont {Zhou}, \citenamefont {Wang},
  \citenamefont {Zhang}, \citenamefont {Weng}, \citenamefont {Prabhakaran},
  \citenamefont {Mo}, \citenamefont {Dudin}, \citenamefont {Kim}, \citenamefont
  {Hoesch}, \citenamefont {Fang}, \citenamefont {Dai}, \citenamefont {Shen},
  \citenamefont {Feng}, \citenamefont {Hussain},\ and\ \citenamefont
  {Chen}}]{Liu2014}%
  \BibitemOpen
  \bibfield  {author} {\bibinfo {author} {\bibfnamefont {Z.~K.}\ \bibnamefont
  {Liu}}, \bibinfo {author} {\bibfnamefont {J.}~\bibnamefont {Jiang}}, \bibinfo
  {author} {\bibfnamefont {B.}~\bibnamefont {Zhou}}, \bibinfo {author}
  {\bibfnamefont {Z.~J.}\ \bibnamefont {Wang}}, \bibinfo {author}
  {\bibfnamefont {Y.}~\bibnamefont {Zhang}}, \bibinfo {author} {\bibfnamefont
  {H.~M.}\ \bibnamefont {Weng}}, \bibinfo {author} {\bibfnamefont
  {D.}~\bibnamefont {Prabhakaran}}, \bibinfo {author} {\bibfnamefont
  {H.}~\bibnamefont {Mo}, \bibfnamefont {S-K.and~Peng}}, \bibinfo {author}
  {\bibfnamefont {P.}~\bibnamefont {Dudin}}, \bibinfo {author} {\bibfnamefont
  {T.}~\bibnamefont {Kim}}, \bibinfo {author} {\bibfnamefont {M.}~\bibnamefont
  {Hoesch}}, \bibinfo {author} {\bibfnamefont {Z.}~\bibnamefont {Fang}},
  \bibinfo {author} {\bibfnamefont {X.}~\bibnamefont {Dai}}, \bibinfo {author}
  {\bibfnamefont {Z.~X.}\ \bibnamefont {Shen}}, \bibinfo {author}
  {\bibfnamefont {D.~L.}\ \bibnamefont {Feng}}, \bibinfo {author}
  {\bibfnamefont {Z.}~\bibnamefont {Hussain}},\ and\ \bibinfo {author}
  {\bibfnamefont {Y.~L.}\ \bibnamefont {Chen}},\ }\href
  {https://doi.org/10.1038/nmat3990} {\bibfield  {journal} {\bibinfo  {journal}
  {Nat. Mater.}\ }\textbf {\bibinfo {volume} {13}},\ \bibinfo {pages} {677}
  (\bibinfo {year} {2014})}\BibitemShut {NoStop}%
\bibitem [{\citenamefont {Liang}\ \emph {et~al.}(2015)\citenamefont {Liang},
  \citenamefont {Gibson}, \citenamefont {Ali}, \citenamefont {Liu},
  \citenamefont {Cava},\ and\ \citenamefont {Ong}}]{Liang2015}%
  \BibitemOpen
  \bibfield  {author} {\bibinfo {author} {\bibfnamefont {T.}~\bibnamefont
  {Liang}}, \bibinfo {author} {\bibfnamefont {Q.}~\bibnamefont {Gibson}},
  \bibinfo {author} {\bibfnamefont {M.~N.}\ \bibnamefont {Ali}}, \bibinfo
  {author} {\bibfnamefont {M.}~\bibnamefont {Liu}}, \bibinfo {author}
  {\bibfnamefont {R.~J.}\ \bibnamefont {Cava}},\ and\ \bibinfo {author}
  {\bibfnamefont {N.~P.}\ \bibnamefont {Ong}},\ }\href
  {https://doi.org/10.1038/nmat4143} {\bibfield  {journal} {\bibinfo  {journal}
  {Nat. Mater.}\ }\textbf {\bibinfo {volume} {14}},\ \bibinfo {pages} {280}
  (\bibinfo {year} {2015})}\BibitemShut {NoStop}%
\bibitem [{\citenamefont {Neupane}\ \emph {et~al.}(2014)\citenamefont
  {Neupane}, \citenamefont {Xu}, \citenamefont {Sankar}, \citenamefont
  {Alidoust}, \citenamefont {Bian}, \citenamefont {Liu}, \citenamefont
  {Belopolski}, \citenamefont {Chang}, \citenamefont {Jeng}, \citenamefont
  {Lin}, \citenamefont {Bansil}, \citenamefont {Chou},\ and\ \citenamefont
  {Hasan}}]{Neupane2014}%
  \BibitemOpen
  \bibfield  {author} {\bibinfo {author} {\bibfnamefont {M.}~\bibnamefont
  {Neupane}}, \bibinfo {author} {\bibfnamefont {S.-Y.}\ \bibnamefont {Xu}},
  \bibinfo {author} {\bibfnamefont {R.}~\bibnamefont {Sankar}}, \bibinfo
  {author} {\bibfnamefont {N.}~\bibnamefont {Alidoust}}, \bibinfo {author}
  {\bibfnamefont {G.}~\bibnamefont {Bian}}, \bibinfo {author} {\bibfnamefont
  {C.}~\bibnamefont {Liu}}, \bibinfo {author} {\bibfnamefont {I.}~\bibnamefont
  {Belopolski}}, \bibinfo {author} {\bibfnamefont {T.-R.}\ \bibnamefont
  {Chang}}, \bibinfo {author} {\bibfnamefont {H.-T.}\ \bibnamefont {Jeng}},
  \bibinfo {author} {\bibfnamefont {H.}~\bibnamefont {Lin}}, \bibinfo {author}
  {\bibfnamefont {A.}~\bibnamefont {Bansil}}, \bibinfo {author} {\bibfnamefont
  {F.}~\bibnamefont {Chou}},\ and\ \bibinfo {author} {\bibfnamefont {M.~Z.}\
  \bibnamefont {Hasan}},\ }\href {https://doi.org/10.1038/ncomms4786}
  {\bibfield  {journal} {\bibinfo  {journal} {Nat. Commun.}\ }\textbf {\bibinfo
  {volume} {5}},\ \bibinfo {pages} {3786} (\bibinfo {year} {2014})}\BibitemShut
  {NoStop}%
\bibitem [{\citenamefont {Wang}\ \emph {et~al.}(2013)\citenamefont {Wang},
  \citenamefont {Weng}, \citenamefont {Wu}, \citenamefont {Dai},\ and\
  \citenamefont {Fang}}]{Wang2013}%
  \BibitemOpen
  \bibfield  {author} {\bibinfo {author} {\bibfnamefont {Z.}~\bibnamefont
  {Wang}}, \bibinfo {author} {\bibfnamefont {H.}~\bibnamefont {Weng}}, \bibinfo
  {author} {\bibfnamefont {Q.}~\bibnamefont {Wu}}, \bibinfo {author}
  {\bibfnamefont {X.}~\bibnamefont {Dai}},\ and\ \bibinfo {author}
  {\bibfnamefont {Z.}~\bibnamefont {Fang}},\ }\href
  {https://doi.org/10.1103/PhysRevB.88.125427} {\bibfield  {journal} {\bibinfo
  {journal} {Phys. Rev. B}\ }\textbf {\bibinfo {volume} {88}},\ \bibinfo
  {pages} {125427} (\bibinfo {year} {2013})}\BibitemShut {NoStop}%
\bibitem [{\citenamefont {Yang}\ and\ \citenamefont
  {Nagaosa}(2014)}]{Yang2014}%
  \BibitemOpen
  \bibfield  {author} {\bibinfo {author} {\bibfnamefont {B.-J.}\ \bibnamefont
  {Yang}}\ and\ \bibinfo {author} {\bibfnamefont {N.}~\bibnamefont {Nagaosa}},\
  }\href {https://doi.org/10.1038/ncomms5898} {\bibfield  {journal} {\bibinfo
  {journal} {Nat. Commun.}\ }\textbf {\bibinfo {volume} {5}},\ \bibinfo {pages}
  {4898} (\bibinfo {year} {2014})}\BibitemShut {NoStop}%
\bibitem [{\citenamefont {Wehling}\ \emph {et~al.}(2014)\citenamefont
  {Wehling}, \citenamefont {Black-Schaffer},\ and\ \citenamefont
  {Balatsky}}]{Wehling2014}%
  \BibitemOpen
  \bibfield  {author} {\bibinfo {author} {\bibfnamefont {T.~O.}\ \bibnamefont
  {Wehling}}, \bibinfo {author} {\bibfnamefont {A.~M.}\ \bibnamefont
  {Black-Schaffer}},\ and\ \bibinfo {author} {\bibfnamefont {A.~V.}\
  \bibnamefont {Balatsky}},\ }\href
  {https://doi.org/10.1080/00018732.2014.927109} {\bibfield  {journal}
  {\bibinfo  {journal} {Adv. Phys.}\ }\textbf {\bibinfo {volume} {63}},\
  \bibinfo {pages} {1} (\bibinfo {year} {2014})}\BibitemShut {NoStop}%
\bibitem [{\citenamefont {Schumann}\ \emph {et~al.}(2016)\citenamefont
  {Schumann}, \citenamefont {Goyal}, \citenamefont {Kim},\ and\ \citenamefont
  {Stemmer}}]{schumann2016molecular}%
  \BibitemOpen
  \bibfield  {author} {\bibinfo {author} {\bibfnamefont {T.}~\bibnamefont
  {Schumann}}, \bibinfo {author} {\bibfnamefont {M.}~\bibnamefont {Goyal}},
  \bibinfo {author} {\bibfnamefont {H.}~\bibnamefont {Kim}},\ and\ \bibinfo
  {author} {\bibfnamefont {S.}~\bibnamefont {Stemmer}},\ }\href
  {https://aip.scitation.org/doi/10.1063/1.4972999} {\bibfield  {journal}
  {\bibinfo  {journal} {APL Mater.}\ }\textbf {\bibinfo {volume} {4}},\
  \bibinfo {pages} {126110} (\bibinfo {year} {2016})}\BibitemShut {NoStop}%
\bibitem [{\citenamefont {Uchida}\ \emph {et~al.}(2017)\citenamefont {Uchida},
  \citenamefont {Nakazawa}, \citenamefont {Nishihaya}, \citenamefont {Akiba},
  \citenamefont {Kriener}, \citenamefont {Kozuka}, \citenamefont {Miyake},
  \citenamefont {Taguchi}, \citenamefont {Tokunaga}, \citenamefont {Nagaosa},
  \citenamefont {Tokura},\ and\ \citenamefont {Kawasaki}}]{uchida2017quantum}%
  \BibitemOpen
  \bibfield  {author} {\bibinfo {author} {\bibfnamefont {M.}~\bibnamefont
  {Uchida}}, \bibinfo {author} {\bibfnamefont {Y.}~\bibnamefont {Nakazawa}},
  \bibinfo {author} {\bibfnamefont {S.}~\bibnamefont {Nishihaya}}, \bibinfo
  {author} {\bibfnamefont {K.}~\bibnamefont {Akiba}}, \bibinfo {author}
  {\bibfnamefont {M.}~\bibnamefont {Kriener}}, \bibinfo {author} {\bibfnamefont
  {Y.}~\bibnamefont {Kozuka}}, \bibinfo {author} {\bibfnamefont
  {A.}~\bibnamefont {Miyake}}, \bibinfo {author} {\bibfnamefont
  {Y.}~\bibnamefont {Taguchi}}, \bibinfo {author} {\bibfnamefont
  {M.}~\bibnamefont {Tokunaga}}, \bibinfo {author} {\bibfnamefont
  {N.}~\bibnamefont {Nagaosa}}, \bibinfo {author} {\bibfnamefont
  {Y.}~\bibnamefont {Tokura}},\ and\ \bibinfo {author} {\bibfnamefont
  {M.}~\bibnamefont {Kawasaki}},\ }\href
  {https://www.nature.com/articles/s41467-017-02423-1} {\bibfield  {journal}
  {\bibinfo  {journal} {Nat. Commun.}\ }\textbf {\bibinfo {volume} {8}},\
  \bibinfo {pages} {2274} (\bibinfo {year} {2017})}\BibitemShut {NoStop}%
\bibitem [{\citenamefont {Schumann}\ \emph {et~al.}(2018)\citenamefont
  {Schumann}, \citenamefont {Galletti}, \citenamefont {Kealhofer},
  \citenamefont {Kim}, \citenamefont {Goyal},\ and\ \citenamefont
  {Stemmer}}]{schumann2018observation}%
  \BibitemOpen
  \bibfield  {author} {\bibinfo {author} {\bibfnamefont {T.}~\bibnamefont
  {Schumann}}, \bibinfo {author} {\bibfnamefont {L.}~\bibnamefont {Galletti}},
  \bibinfo {author} {\bibfnamefont {D.~A.}\ \bibnamefont {Kealhofer}}, \bibinfo
  {author} {\bibfnamefont {H.}~\bibnamefont {Kim}}, \bibinfo {author}
  {\bibfnamefont {M.}~\bibnamefont {Goyal}},\ and\ \bibinfo {author}
  {\bibfnamefont {S.}~\bibnamefont {Stemmer}},\ }\href
  {https://journals.aps.org/prl/abstract/10.1103/PhysRevLett.120.016801}
  {\bibfield  {journal} {\bibinfo  {journal} {Phys. Rev. Lett.}\ }\textbf
  {\bibinfo {volume} {120}},\ \bibinfo {pages} {016801} (\bibinfo {year}
  {2018})}\BibitemShut {NoStop}%
\bibitem [{\citenamefont {Galletti}\ \emph {et~al.}(2019)\citenamefont
  {Galletti}, \citenamefont {Schumann}, \citenamefont {Kealhofer},
  \citenamefont {Goyal},\ and\ \citenamefont
  {Stemmer}}]{Galletti_PhysRevB.99.201401}%
  \BibitemOpen
  \bibfield  {author} {\bibinfo {author} {\bibfnamefont {L.}~\bibnamefont
  {Galletti}}, \bibinfo {author} {\bibfnamefont {T.}~\bibnamefont {Schumann}},
  \bibinfo {author} {\bibfnamefont {D.~A.}\ \bibnamefont {Kealhofer}}, \bibinfo
  {author} {\bibfnamefont {M.}~\bibnamefont {Goyal}},\ and\ \bibinfo {author}
  {\bibfnamefont {S.}~\bibnamefont {Stemmer}},\ }\href
  {https://doi.org/10.1103/PhysRevB.99.201401} {\bibfield  {journal} {\bibinfo
  {journal} {Phys. Rev. B}\ }\textbf {\bibinfo {volume} {99}},\ \bibinfo
  {pages} {201401(R)} (\bibinfo {year} {2019})}\BibitemShut {NoStop}%
\bibitem [{\citenamefont {Xiao}\ \emph
  {et~al.}(2022{\natexlab{b}})\citenamefont {Xiao}, \citenamefont {Zhang},
  \citenamefont {Chamorro}, \citenamefont {Kim}, \citenamefont {McQueen},
  \citenamefont {Vanderbilt}, \citenamefont {Kayyalha}, \citenamefont {Li},\
  and\ \citenamefont {Samarth}}]{Xiao_PhysRevB.106.L201101}%
  \BibitemOpen
  \bibfield  {author} {\bibinfo {author} {\bibfnamefont {R.}~\bibnamefont
  {Xiao}}, \bibinfo {author} {\bibfnamefont {J.}~\bibnamefont {Zhang}},
  \bibinfo {author} {\bibfnamefont {J.}~\bibnamefont {Chamorro}}, \bibinfo
  {author} {\bibfnamefont {J.}~\bibnamefont {Kim}}, \bibinfo {author}
  {\bibfnamefont {T.~M.}\ \bibnamefont {McQueen}}, \bibinfo {author}
  {\bibfnamefont {D.}~\bibnamefont {Vanderbilt}}, \bibinfo {author}
  {\bibfnamefont {M.}~\bibnamefont {Kayyalha}}, \bibinfo {author}
  {\bibfnamefont {Y.}~\bibnamefont {Li}},\ and\ \bibinfo {author}
  {\bibfnamefont {N.}~\bibnamefont {Samarth}},\ }\href
  {https://doi.org/10.1103/PhysRevB.106.L201101} {\bibfield  {journal}
  {\bibinfo  {journal} {Phys. Rev. B}\ }\textbf {\bibinfo {volume} {106}},\
  \bibinfo {pages} {L201101} (\bibinfo {year}
  {2022}{\natexlab{b}})}\BibitemShut {NoStop}%
\bibitem [{\citenamefont {Shoron}\ \emph {et~al.}(2021)\citenamefont {Shoron},
  \citenamefont {Kealhofer}, \citenamefont {Goyal}, \citenamefont {Schumann},\
  and\ \citenamefont {Stemmer}}]{Shoron_2021}%
  \BibitemOpen
  \bibfield  {author} {\bibinfo {author} {\bibfnamefont {O.~F.}\ \bibnamefont
  {Shoron}}, \bibinfo {author} {\bibfnamefont {D.~A.}\ \bibnamefont
  {Kealhofer}}, \bibinfo {author} {\bibfnamefont {M.}~\bibnamefont {Goyal}},
  \bibinfo {author} {\bibfnamefont {A.}~\bibnamefont {Schumann}, \bibfnamefont
  {Timo~Burkov}},\ and\ \bibinfo {author} {\bibfnamefont {S.}~\bibnamefont
  {Stemmer}},\ }\href
  {https://pubs.aip.org/aip/apl/article/119/17/171907/40638/Detecting-topological-phase-transitions-in-cadmium}
  {\bibfield  {journal} {\bibinfo  {journal} {Appl. Phys. Lett.}\ }\textbf
  {\bibinfo {volume} {119}},\ \bibinfo {pages} {171907} (\bibinfo {year}
  {2021})}\BibitemShut {NoStop}%
\bibitem [{\citenamefont {Yanez}\ \emph {et~al.}(2021)\citenamefont {Yanez},
  \citenamefont {Ou}, \citenamefont {Xiao}, \citenamefont {Koo}, \citenamefont
  {Held}, \citenamefont {Ghosh}, \citenamefont {Rable}, \citenamefont
  {Pillsbury}, \citenamefont {Delgado}, \citenamefont {Yang}, \citenamefont
  {Chamorro}, \citenamefont {Grutter}, \citenamefont {Quarterman},
  \citenamefont {Richardella}, \citenamefont {Sengupta}, \citenamefont
  {McQueen}, \citenamefont {Borchers}, \citenamefont {Mkhoyan}, \citenamefont
  {Yan},\ and\ \citenamefont {Samarth}}]{Yanez_PhysRevApplied.16.054031}%
  \BibitemOpen
  \bibfield  {author} {\bibinfo {author} {\bibfnamefont {W.}~\bibnamefont
  {Yanez}}, \bibinfo {author} {\bibfnamefont {Y.}~\bibnamefont {Ou}}, \bibinfo
  {author} {\bibfnamefont {R.}~\bibnamefont {Xiao}}, \bibinfo {author}
  {\bibfnamefont {J.}~\bibnamefont {Koo}}, \bibinfo {author} {\bibfnamefont
  {J.~T.}\ \bibnamefont {Held}}, \bibinfo {author} {\bibfnamefont
  {S.}~\bibnamefont {Ghosh}}, \bibinfo {author} {\bibfnamefont
  {J.}~\bibnamefont {Rable}}, \bibinfo {author} {\bibfnamefont
  {T.}~\bibnamefont {Pillsbury}}, \bibinfo {author} {\bibfnamefont {E.~G.}\
  \bibnamefont {Delgado}}, \bibinfo {author} {\bibfnamefont {K.}~\bibnamefont
  {Yang}}, \bibinfo {author} {\bibfnamefont {J.}~\bibnamefont {Chamorro}},
  \bibinfo {author} {\bibfnamefont {A.~J.}\ \bibnamefont {Grutter}}, \bibinfo
  {author} {\bibfnamefont {P.}~\bibnamefont {Quarterman}}, \bibinfo {author}
  {\bibfnamefont {A.}~\bibnamefont {Richardella}}, \bibinfo {author}
  {\bibfnamefont {A.}~\bibnamefont {Sengupta}}, \bibinfo {author}
  {\bibfnamefont {T.}~\bibnamefont {McQueen}}, \bibinfo {author} {\bibfnamefont
  {J.~A.}\ \bibnamefont {Borchers}}, \bibinfo {author} {\bibfnamefont {K.~A.}\
  \bibnamefont {Mkhoyan}}, \bibinfo {author} {\bibfnamefont {B.}~\bibnamefont
  {Yan}},\ and\ \bibinfo {author} {\bibfnamefont {N.}~\bibnamefont {Samarth}},\
  }\href {https://doi.org/10.1103/PhysRevApplied.16.054031} {\bibfield
  {journal} {\bibinfo  {journal} {Phys. Rev. Applied}\ }\textbf {\bibinfo
  {volume} {16}},\ \bibinfo {pages} {054031} (\bibinfo {year}
  {2021})}\BibitemShut {NoStop}%
\bibitem [{\citenamefont {Ali}\ \emph {et~al.}(2014)\citenamefont {Ali},
  \citenamefont {Gibson}, \citenamefont {Jeon}, \citenamefont {Zhou},
  \citenamefont {Yazdani},\ and\ \citenamefont {Cava}}]{Cava}%
  \BibitemOpen
  \bibfield  {author} {\bibinfo {author} {\bibfnamefont {M.~N.}\ \bibnamefont
  {Ali}}, \bibinfo {author} {\bibfnamefont {Q.}~\bibnamefont {Gibson}},
  \bibinfo {author} {\bibfnamefont {S.}~\bibnamefont {Jeon}}, \bibinfo {author}
  {\bibfnamefont {B.~B.}\ \bibnamefont {Zhou}}, \bibinfo {author}
  {\bibfnamefont {A.}~\bibnamefont {Yazdani}},\ and\ \bibinfo {author}
  {\bibfnamefont {R.~J.}\ \bibnamefont {Cava}},\ }\href
  {https://pubs.acs.org/doi/abs/10.1021/ic403163d} {\bibfield  {journal}
  {\bibinfo  {journal} {Inorg. Chem.}\ }\textbf {\bibinfo {volume} {53}},\
  \bibinfo {pages} {4062} (\bibinfo {year} {2014})}\BibitemShut {NoStop}%
\bibitem [{\citenamefont {Steigmann}\ and\ \citenamefont
  {Goodyear}(1968)}]{Steigmann}%
  \BibitemOpen
  \bibfield  {author} {\bibinfo {author} {\bibfnamefont {G.~A.}\ \bibnamefont
  {Steigmann}}\ and\ \bibinfo {author} {\bibfnamefont {J.}~\bibnamefont
  {Goodyear}},\ }\href {https://doi.org/10.1107/S0567740868003705} {\bibfield
  {journal} {\bibinfo  {journal} {Acta Crystallogr. Sec. B}\ }\textbf {\bibinfo
  {volume} {24}},\ \bibinfo {pages} {1062} (\bibinfo {year}
  {1968})}\BibitemShut {NoStop}%
\bibitem [{\citenamefont {Uchida}\ \emph {et~al.}(2019)\citenamefont {Uchida},
  \citenamefont {Koretsune}, \citenamefont {Sato}, \citenamefont {Kriener},
  \citenamefont {Nakazawa}, \citenamefont {Nishihaya}, \citenamefont {Taguchi},
  \citenamefont {Arita},\ and\ \citenamefont
  {Kawasaki}}]{Uchida_PhysRevB.100.245148}%
  \BibitemOpen
  \bibfield  {author} {\bibinfo {author} {\bibfnamefont {M.}~\bibnamefont
  {Uchida}}, \bibinfo {author} {\bibfnamefont {T.}~\bibnamefont {Koretsune}},
  \bibinfo {author} {\bibfnamefont {S.}~\bibnamefont {Sato}}, \bibinfo {author}
  {\bibfnamefont {M.}~\bibnamefont {Kriener}}, \bibinfo {author} {\bibfnamefont
  {Y.}~\bibnamefont {Nakazawa}}, \bibinfo {author} {\bibfnamefont
  {S.}~\bibnamefont {Nishihaya}}, \bibinfo {author} {\bibfnamefont
  {Y.}~\bibnamefont {Taguchi}}, \bibinfo {author} {\bibfnamefont
  {R.}~\bibnamefont {Arita}},\ and\ \bibinfo {author} {\bibfnamefont
  {M.}~\bibnamefont {Kawasaki}},\ }\href
  {https://doi.org/10.1103/PhysRevB.100.245148} {\bibfield  {journal} {\bibinfo
   {journal} {Phys. Rev. B}\ }\textbf {\bibinfo {volume} {100}},\ \bibinfo
  {pages} {245148} (\bibinfo {year} {2019})}\BibitemShut {NoStop}%
\bibitem [{\citenamefont {Zutic}\ \emph {et~al.}(2019)\citenamefont {Zutic},
  \citenamefont {Matos-Abiague}, \citenamefont {Scharf}, \citenamefont {Dery},\
  and\ \citenamefont {Belashchenko}}]{Zutic_MT}%
  \BibitemOpen
  \bibfield  {author} {\bibinfo {author} {\bibfnamefont {I.}~\bibnamefont
  {Zutic}}, \bibinfo {author} {\bibfnamefont {A.}~\bibnamefont
  {Matos-Abiague}}, \bibinfo {author} {\bibfnamefont {B.}~\bibnamefont
  {Scharf}}, \bibinfo {author} {\bibfnamefont {H.}~\bibnamefont {Dery}},\ and\
  \bibinfo {author} {\bibfnamefont {K.}~\bibnamefont {Belashchenko}},\ }\href
  {https://doi.org/10.1016/j.mattod.2018.05.003} {\bibfield  {journal}
  {\bibinfo  {journal} {Materials Today}\ }\textbf {\bibinfo {volume} {22}},\
  \bibinfo {pages} {85} (\bibinfo {year} {2019})}\BibitemShut {NoStop}%
\bibitem [{\citenamefont {Schumann}\ \emph {et~al.}(2017)\citenamefont
  {Schumann}, \citenamefont {Goyal}, \citenamefont {Kealhofer},\ and\
  \citenamefont {Stemmer}}]{Schumann2017}%
  \BibitemOpen
  \bibfield  {author} {\bibinfo {author} {\bibfnamefont {T.}~\bibnamefont
  {Schumann}}, \bibinfo {author} {\bibfnamefont {M.}~\bibnamefont {Goyal}},
  \bibinfo {author} {\bibfnamefont {D.~A.}\ \bibnamefont {Kealhofer}},\ and\
  \bibinfo {author} {\bibfnamefont {S.}~\bibnamefont {Stemmer}},\ }\href
  {https://doi.org/10.1103/PhysRevB.95.241113} {\bibfield  {journal} {\bibinfo
  {journal} {Phys. Rev. B}\ }\textbf {\bibinfo {volume} {95}},\ \bibinfo
  {pages} {241113} (\bibinfo {year} {2017})}\BibitemShut {NoStop}%
\bibitem [{\citenamefont {Matsukura}\ \emph {et~al.}(2000)\citenamefont
  {Matsukura}, \citenamefont {Abe},\ and\ \citenamefont
  {Ohno}}]{Matsukara_JAP}%
  \BibitemOpen
  \bibfield  {author} {\bibinfo {author} {\bibfnamefont {F.}~\bibnamefont
  {Matsukura}}, \bibinfo {author} {\bibfnamefont {E.}~\bibnamefont {Abe}},\
  and\ \bibinfo {author} {\bibfnamefont {H.}~\bibnamefont {Ohno}},\ }\href
  {https://aip.scitation.org/doi/10.1063/1.372732} {\bibfield  {journal}
  {\bibinfo  {journal} {J. Appl. Phys.}\ }\textbf {\bibinfo {volume} {87}},\
  \bibinfo {pages} {6442} (\bibinfo {year} {2000})}\BibitemShut {NoStop}%
\bibitem [{\citenamefont {McCombe}\ \emph {et~al.}(2003)\citenamefont
  {McCombe}, \citenamefont {Na}, \citenamefont {Chen}, \citenamefont {Cheon},
  \citenamefont {Wang}, \citenamefont {Luo}, \citenamefont {Liu}, \citenamefont
  {Sasaki}, \citenamefont {Wojtowicz}, \citenamefont {Furdyna}, \citenamefont
  {Potashnik},\ and\ \citenamefont {Schiffer}}]{MCCOMBE200390}%
  \BibitemOpen
  \bibfield  {author} {\bibinfo {author} {\bibfnamefont {B.}~\bibnamefont
  {McCombe}}, \bibinfo {author} {\bibfnamefont {M.}~\bibnamefont {Na}},
  \bibinfo {author} {\bibfnamefont {X.}~\bibnamefont {Chen}}, \bibinfo {author}
  {\bibfnamefont {M.}~\bibnamefont {Cheon}}, \bibinfo {author} {\bibfnamefont
  {S.}~\bibnamefont {Wang}}, \bibinfo {author} {\bibfnamefont {H.}~\bibnamefont
  {Luo}}, \bibinfo {author} {\bibfnamefont {X.}~\bibnamefont {Liu}}, \bibinfo
  {author} {\bibfnamefont {Y.}~\bibnamefont {Sasaki}}, \bibinfo {author}
  {\bibfnamefont {T.}~\bibnamefont {Wojtowicz}}, \bibinfo {author}
  {\bibfnamefont {J.}~\bibnamefont {Furdyna}}, \bibinfo {author} {\bibfnamefont
  {S.}~\bibnamefont {Potashnik}},\ and\ \bibinfo {author} {\bibfnamefont
  {P.}~\bibnamefont {Schiffer}},\ }\href
  {https://doi.org/https://doi.org/10.1016/S1386-9477(02)00594-5} {\bibfield
  {journal} {\bibinfo  {journal} {Physica E: Low-dimensional Systems and
  Nanostructures}\ }\textbf {\bibinfo {volume} {16}},\ \bibinfo {pages} {90}
  (\bibinfo {year} {2003})}\BibitemShut {NoStop}%
\bibitem [{\citenamefont {Riddiford}\ \emph
  {et~al.}(2022{\natexlab{b}})\citenamefont {Riddiford}, \citenamefont
  {Grutter}, \citenamefont {Pillsbury}, \citenamefont {Stanley}, \citenamefont
  {Reifsnyder~Hickey}, \citenamefont {Li}, \citenamefont {Alem}, \citenamefont
  {Samarth},\ and\ \citenamefont {Suzuki}}]{PhysRevLett.128.126802}%
  \BibitemOpen
  \bibfield  {author} {\bibinfo {author} {\bibfnamefont {L.~J.}\ \bibnamefont
  {Riddiford}}, \bibinfo {author} {\bibfnamefont {A.~J.}\ \bibnamefont
  {Grutter}}, \bibinfo {author} {\bibfnamefont {T.}~\bibnamefont {Pillsbury}},
  \bibinfo {author} {\bibfnamefont {M.}~\bibnamefont {Stanley}}, \bibinfo
  {author} {\bibfnamefont {D.}~\bibnamefont {Reifsnyder~Hickey}}, \bibinfo
  {author} {\bibfnamefont {P.}~\bibnamefont {Li}}, \bibinfo {author}
  {\bibfnamefont {N.}~\bibnamefont {Alem}}, \bibinfo {author} {\bibfnamefont
  {N.}~\bibnamefont {Samarth}},\ and\ \bibinfo {author} {\bibfnamefont
  {Y.}~\bibnamefont {Suzuki}},\ }\href
  {https://doi.org/10.1103/PhysRevLett.128.126802} {\bibfield  {journal}
  {\bibinfo  {journal} {Phys. Rev. Lett.}\ }\textbf {\bibinfo {volume} {128}},\
  \bibinfo {pages} {126802} (\bibinfo {year} {2022}{\natexlab{b}})}\BibitemShut
  {NoStop}%
\bibitem [{\citenamefont {Lawniczak-Jablonska}\ \emph
  {et~al.}(2011)\citenamefont {Lawniczak-Jablonska}, \citenamefont {Wolska},
  \citenamefont {Klepka}, \citenamefont {Kret}, \citenamefont {Gosk},
  \citenamefont {Twardowski}, \citenamefont {Wasik}, \citenamefont
  {Kwiatkowski}, \citenamefont {Kurowska}, \citenamefont {Kowalski},\ and\
  \citenamefont {Sadowski}}]{Jabolonska_doi:10.1063/1.3562171}%
  \BibitemOpen
  \bibfield  {author} {\bibinfo {author} {\bibfnamefont {K.}~\bibnamefont
  {Lawniczak-Jablonska}}, \bibinfo {author} {\bibfnamefont {A.}~\bibnamefont
  {Wolska}}, \bibinfo {author} {\bibfnamefont {M.~T.}\ \bibnamefont {Klepka}},
  \bibinfo {author} {\bibfnamefont {S.}~\bibnamefont {Kret}}, \bibinfo {author}
  {\bibfnamefont {J.}~\bibnamefont {Gosk}}, \bibinfo {author} {\bibfnamefont
  {A.}~\bibnamefont {Twardowski}}, \bibinfo {author} {\bibfnamefont
  {D.}~\bibnamefont {Wasik}}, \bibinfo {author} {\bibfnamefont
  {A.}~\bibnamefont {Kwiatkowski}}, \bibinfo {author} {\bibfnamefont
  {B.}~\bibnamefont {Kurowska}}, \bibinfo {author} {\bibfnamefont {B.~J.}\
  \bibnamefont {Kowalski}},\ and\ \bibinfo {author} {\bibfnamefont
  {J.}~\bibnamefont {Sadowski}},\ }\href {https://doi.org/10.1063/1.3562171}
  {\bibfield  {journal} {\bibinfo  {journal} {J. Appl. Phys.}\ }\textbf
  {\bibinfo {volume} {109}},\ \bibinfo {pages} {074308} (\bibinfo {year}
  {2011})}\BibitemShut {NoStop}%
\bibitem [{Mit()}]{Mitra_supp}%
  \BibitemOpen
  \href@noop {} {}\bibinfo {note} {See Supplemental Material at [URL will be
  inserted by publisher] for additional PNR, TEM, magnetometry and transport
  data}\BibitemShut {NoStop}%
\bibitem [{\citenamefont {Borisenko}\ \emph {et~al.}(2014)\citenamefont
  {Borisenko}, \citenamefont {Gibson}, \citenamefont {Evtushinsky},
  \citenamefont {Zabolotnyy}, \citenamefont {B\"uchner},\ and\ \citenamefont
  {Cava}}]{Borisenko2014}%
  \BibitemOpen
  \bibfield  {author} {\bibinfo {author} {\bibfnamefont {S.}~\bibnamefont
  {Borisenko}}, \bibinfo {author} {\bibfnamefont {Q.}~\bibnamefont {Gibson}},
  \bibinfo {author} {\bibfnamefont {D.}~\bibnamefont {Evtushinsky}}, \bibinfo
  {author} {\bibfnamefont {V.}~\bibnamefont {Zabolotnyy}}, \bibinfo {author}
  {\bibfnamefont {B.}~\bibnamefont {B\"uchner}},\ and\ \bibinfo {author}
  {\bibfnamefont {R.~J.}\ \bibnamefont {Cava}},\ }\href
  {https://doi.org/10.1103/PhysRevLett.113.027603} {\bibfield  {journal}
  {\bibinfo  {journal} {Phys. Rev. Lett.}\ }\textbf {\bibinfo {volume} {113}},\
  \bibinfo {pages} {027603} (\bibinfo {year} {2014})}\BibitemShut {NoStop}%
\bibitem [{\citenamefont {Yi}\ \emph {et~al.}(2014)\citenamefont {Yi},
  \citenamefont {Wang}, \citenamefont {Chen}, \citenamefont {Shi},
  \citenamefont {Feng}, \citenamefont {Liang}, \citenamefont {Xie},
  \citenamefont {He}, \citenamefont {He}, \citenamefont {Peng}, \citenamefont
  {Liu}, \citenamefont {Liu}, \citenamefont {Zhao}, \citenamefont {Liu},
  \citenamefont {Dong}, \citenamefont {Zhang}, \citenamefont {Nakatake},
  \citenamefont {Arita}, \citenamefont {Shimada}, \citenamefont {Namatame},
  \citenamefont {Taniguchi}, \citenamefont {Xu}, \citenamefont {Chen},
  \citenamefont {Dai}, \citenamefont {Fang},\ and\ \citenamefont
  {Zhou}}]{Yi2014}%
  \BibitemOpen
  \bibfield  {author} {\bibinfo {author} {\bibfnamefont {H.}~\bibnamefont
  {Yi}}, \bibinfo {author} {\bibfnamefont {Z.}~\bibnamefont {Wang}}, \bibinfo
  {author} {\bibfnamefont {C.}~\bibnamefont {Chen}}, \bibinfo {author}
  {\bibfnamefont {Y.}~\bibnamefont {Shi}}, \bibinfo {author} {\bibfnamefont
  {Y.}~\bibnamefont {Feng}}, \bibinfo {author} {\bibfnamefont {A.}~\bibnamefont
  {Liang}}, \bibinfo {author} {\bibfnamefont {Z.}~\bibnamefont {Xie}}, \bibinfo
  {author} {\bibfnamefont {S.}~\bibnamefont {He}}, \bibinfo {author}
  {\bibfnamefont {J.}~\bibnamefont {He}}, \bibinfo {author} {\bibfnamefont
  {Y.}~\bibnamefont {Peng}}, \bibinfo {author} {\bibfnamefont {X.}~\bibnamefont
  {Liu}}, \bibinfo {author} {\bibfnamefont {Y.}~\bibnamefont {Liu}}, \bibinfo
  {author} {\bibfnamefont {L.}~\bibnamefont {Zhao}}, \bibinfo {author}
  {\bibfnamefont {G.}~\bibnamefont {Liu}}, \bibinfo {author} {\bibfnamefont
  {X.}~\bibnamefont {Dong}}, \bibinfo {author} {\bibfnamefont {J.}~\bibnamefont
  {Zhang}}, \bibinfo {author} {\bibfnamefont {M.}~\bibnamefont {Nakatake}},
  \bibinfo {author} {\bibfnamefont {M.}~\bibnamefont {Arita}}, \bibinfo
  {author} {\bibfnamefont {K.}~\bibnamefont {Shimada}}, \bibinfo {author}
  {\bibfnamefont {H.}~\bibnamefont {Namatame}}, \bibinfo {author}
  {\bibfnamefont {M.}~\bibnamefont {Taniguchi}}, \bibinfo {author}
  {\bibfnamefont {Z.}~\bibnamefont {Xu}}, \bibinfo {author} {\bibfnamefont
  {C.}~\bibnamefont {Chen}}, \bibinfo {author} {\bibfnamefont {X.}~\bibnamefont
  {Dai}}, \bibinfo {author} {\bibfnamefont {Z.}~\bibnamefont {Fang}},\ and\
  \bibinfo {author} {\bibfnamefont {X.~J.}\ \bibnamefont {Zhou}},\ }\href
  {https://doi.org/10.1038/srep06106} {\bibfield  {journal} {\bibinfo
  {journal} {Sci. Rep.}\ }\textbf {\bibinfo {volume} {4}},\ \bibinfo {pages}
  {6106} (\bibinfo {year} {2014})}\BibitemShut {NoStop}%
\bibitem [{\citenamefont {Kisslinger}\ \emph {et~al.}(2017)\citenamefont
  {Kisslinger}, \citenamefont {Ott},\ and\ \citenamefont
  {Weber}}]{Kisslinger2017}%
  \BibitemOpen
  \bibfield  {author} {\bibinfo {author} {\bibfnamefont {F.}~\bibnamefont
  {Kisslinger}}, \bibinfo {author} {\bibfnamefont {C.}~\bibnamefont {Ott}},\
  and\ \bibinfo {author} {\bibfnamefont {H.~B.}\ \bibnamefont {Weber}},\ }\href
  {https://doi.org/10.1103/PhysRevB.95.024204} {\bibfield  {journal} {\bibinfo
  {journal} {Phys. Rev. B}\ }\textbf {\bibinfo {volume} {95}},\ \bibinfo
  {pages} {024204} (\bibinfo {year} {2017})}\BibitemShut {NoStop}%
\bibitem [{\citenamefont {Parish}\ and\ \citenamefont
  {Littlewood}(2003)}]{Parish2003}%
  \BibitemOpen
  \bibfield  {author} {\bibinfo {author} {\bibfnamefont {M.~M.}\ \bibnamefont
  {Parish}}\ and\ \bibinfo {author} {\bibfnamefont {P.~B.}\ \bibnamefont
  {Littlewood}},\ }\href {https://doi.org/10.1038/nature02073} {\bibfield
  {journal} {\bibinfo  {journal} {Nature (London)}\ }\textbf {\bibinfo {volume}
  {426}},\ \bibinfo {pages} {162} (\bibinfo {year} {2003})}\BibitemShut
  {NoStop}%
\bibitem [{\citenamefont {Kozlova}\ \emph {et~al.}(2012)\citenamefont
  {Kozlova}, \citenamefont {Mori}, \citenamefont {Makarovsky}, \citenamefont
  {Eaves}, \citenamefont {Zhuang}, \citenamefont {Krier},\ and\ \citenamefont
  {Patanè}}]{Kozlova2012}%
  \BibitemOpen
  \bibfield  {author} {\bibinfo {author} {\bibfnamefont {N.}~\bibnamefont
  {Kozlova}}, \bibinfo {author} {\bibfnamefont {N.}~\bibnamefont {Mori}},
  \bibinfo {author} {\bibfnamefont {O.}~\bibnamefont {Makarovsky}}, \bibinfo
  {author} {\bibfnamefont {L.}~\bibnamefont {Eaves}}, \bibinfo {author}
  {\bibfnamefont {Q.}~\bibnamefont {Zhuang}}, \bibinfo {author} {\bibfnamefont
  {A.}~\bibnamefont {Krier}},\ and\ \bibinfo {author} {\bibfnamefont
  {A.}~\bibnamefont {Patanè}},\ }\href {https://doi.org/10.1038/ncomms2106}
  {\bibfield  {journal} {\bibinfo  {journal} {Nat. Commun.}\ }\textbf {\bibinfo
  {volume} {3}},\ \bibinfo {pages} {1097} (\bibinfo {year} {2012})}\BibitemShut
  {NoStop}%
\bibitem [{\citenamefont {Jeon}\ \emph {et~al.}(2014)\citenamefont {Jeon},
  \citenamefont {Zhou}, \citenamefont {Gyenis}, \citenamefont {Feldman},
  \citenamefont {Kimchi}, \citenamefont {Potter}, \citenamefont {Gibson},
  \citenamefont {Cava}, \citenamefont {Vishwanath},\ and\ \citenamefont
  {Yazdani}}]{Jeon2014}%
  \BibitemOpen
  \bibfield  {author} {\bibinfo {author} {\bibfnamefont {S.}~\bibnamefont
  {Jeon}}, \bibinfo {author} {\bibfnamefont {B.~B.}\ \bibnamefont {Zhou}},
  \bibinfo {author} {\bibfnamefont {A.}~\bibnamefont {Gyenis}}, \bibinfo
  {author} {\bibfnamefont {B.~E.}\ \bibnamefont {Feldman}}, \bibinfo {author}
  {\bibfnamefont {I.}~\bibnamefont {Kimchi}}, \bibinfo {author} {\bibfnamefont
  {A.~C.}\ \bibnamefont {Potter}}, \bibinfo {author} {\bibfnamefont {Q.~D.}\
  \bibnamefont {Gibson}}, \bibinfo {author} {\bibfnamefont {R.~J.}\
  \bibnamefont {Cava}}, \bibinfo {author} {\bibfnamefont {A.}~\bibnamefont
  {Vishwanath}},\ and\ \bibinfo {author} {\bibfnamefont {A.}~\bibnamefont
  {Yazdani}},\ }\href {https://doi.org/10.1038/nmat4023} {\bibfield  {journal}
  {\bibinfo  {journal} {Nat. Mater.}\ }\textbf {\bibinfo {volume} {13}},\
  \bibinfo {pages} {851} (\bibinfo {year} {2014})}\BibitemShut {NoStop}%
\bibitem [{\citenamefont {Maranville}\ \emph {et~al.}(2018)\citenamefont
  {Maranville}, \citenamefont {Ratcliff~II},\ and\ \citenamefont
  {Kienzle}}]{Maranville:po5131}%
  \BibitemOpen
  \bibfield  {author} {\bibinfo {author} {\bibfnamefont {B.}~\bibnamefont
  {Maranville}}, \bibinfo {author} {\bibfnamefont {W.}~\bibnamefont
  {Ratcliff~II}},\ and\ \bibinfo {author} {\bibfnamefont {P.}~\bibnamefont
  {Kienzle}},\ }\href {https://doi.org/10.1107/S1600576718011974} {\bibfield
  {journal} {\bibinfo  {journal} {Journal of Applied Crystallography}\ }\textbf
  {\bibinfo {volume} {51}},\ \bibinfo {pages} {1500} (\bibinfo {year}
  {2018})}\BibitemShut {NoStop}%
\bibitem [{\citenamefont {Kirby}\ \emph {et~al.}(2012)\citenamefont {Kirby},
  \citenamefont {Kienzle}, \citenamefont {Maranville}, \citenamefont {Berk},
  \citenamefont {Krycka}, \citenamefont {Heinrich},\ and\ \citenamefont
  {Majkrzak}}]{KIRBY201244}%
  \BibitemOpen
  \bibfield  {author} {\bibinfo {author} {\bibfnamefont {B.}~\bibnamefont
  {Kirby}}, \bibinfo {author} {\bibfnamefont {P.}~\bibnamefont {Kienzle}},
  \bibinfo {author} {\bibfnamefont {B.}~\bibnamefont {Maranville}}, \bibinfo
  {author} {\bibfnamefont {N.}~\bibnamefont {Berk}}, \bibinfo {author}
  {\bibfnamefont {J.}~\bibnamefont {Krycka}}, \bibinfo {author} {\bibfnamefont
  {F.}~\bibnamefont {Heinrich}},\ and\ \bibinfo {author} {\bibfnamefont
  {C.}~\bibnamefont {Majkrzak}},\ }\href
  {https://doi.org/https://doi.org/10.1016/j.cocis.2011.11.001} {\bibfield
  {journal} {\bibinfo  {journal} {Current Opinion in Colloid and Interface
  Science}\ }\textbf {\bibinfo {volume} {17}},\ \bibinfo {pages} {44} (\bibinfo
  {year} {2012})}\BibitemShut {NoStop}%
\bibitem [{\citenamefont {Strambini}\ \emph {et~al.}(2017)\citenamefont
  {Strambini}, \citenamefont {Golovach}, \citenamefont {De~Simoni},
  \citenamefont {Moodera}, \citenamefont {Bergeret},\ and\ \citenamefont
  {Giazotto}}]{Strambini_WOS:000416584100003}%
  \BibitemOpen
  \bibfield  {author} {\bibinfo {author} {\bibfnamefont {E.}~\bibnamefont
  {Strambini}}, \bibinfo {author} {\bibfnamefont {V.~N.}\ \bibnamefont
  {Golovach}}, \bibinfo {author} {\bibfnamefont {G.}~\bibnamefont {De~Simoni}},
  \bibinfo {author} {\bibfnamefont {J.~S.}\ \bibnamefont {Moodera}}, \bibinfo
  {author} {\bibfnamefont {F.~S.}\ \bibnamefont {Bergeret}},\ and\ \bibinfo
  {author} {\bibfnamefont {F.}~\bibnamefont {Giazotto}},\ }\href
  {https://doi.org/10.1103/PhysRevMaterials.1.054402} {\bibfield  {journal}
  {\bibinfo  {journal} {Phys. Rev. Mater.}\ }\textbf {\bibinfo {volume} {1}},\
  \bibinfo {pages} {054402} (\bibinfo {year} {2017})}\BibitemShut {NoStop}%
\bibitem [{\citenamefont {Figueroa}\ \emph {et~al.}(2020)\citenamefont
  {Figueroa}, \citenamefont {Bonell}, \citenamefont {Cuxart}, \citenamefont
  {Valvidares}, \citenamefont {Gargiani}, \citenamefont {van~der Laan},
  \citenamefont {Mugarza},\ and\ \citenamefont
  {Valenzuela}}]{Figueroa_PhysRevLett.125.226801}%
  \BibitemOpen
  \bibfield  {author} {\bibinfo {author} {\bibfnamefont {A.~I.}\ \bibnamefont
  {Figueroa}}, \bibinfo {author} {\bibfnamefont {F.}~\bibnamefont {Bonell}},
  \bibinfo {author} {\bibfnamefont {M.~G.}\ \bibnamefont {Cuxart}}, \bibinfo
  {author} {\bibfnamefont {M.}~\bibnamefont {Valvidares}}, \bibinfo {author}
  {\bibfnamefont {P.}~\bibnamefont {Gargiani}}, \bibinfo {author}
  {\bibfnamefont {G.}~\bibnamefont {van~der Laan}}, \bibinfo {author}
  {\bibfnamefont {A.}~\bibnamefont {Mugarza}},\ and\ \bibinfo {author}
  {\bibfnamefont {S.~O.}\ \bibnamefont {Valenzuela}},\ }\href
  {https://doi.org/10.1103/PhysRevLett.125.226801} {\bibfield  {journal}
  {\bibinfo  {journal} {Phys. Rev. Lett.}\ }\textbf {\bibinfo {volume} {125}},\
  \bibinfo {pages} {226801} (\bibinfo {year} {2020})}\BibitemShut {NoStop}%
\bibitem [{\citenamefont {Zhang}\ \emph {et~al.}(2014)\citenamefont {Zhang},
  \citenamefont {Feng}, \citenamefont {Guo}, \citenamefont {Li}, \citenamefont
  {Zhang}, \citenamefont {Ou}, \citenamefont {Feng}, \citenamefont {Wang},
  \citenamefont {Chen}, \citenamefont {He}, \citenamefont {Ma}, \citenamefont
  {Xue},\ and\ \citenamefont {Wang}}]{Zhang2014}%
  \BibitemOpen
  \bibfield  {author} {\bibinfo {author} {\bibfnamefont {Z.}~\bibnamefont
  {Zhang}}, \bibinfo {author} {\bibfnamefont {X.}~\bibnamefont {Feng}},
  \bibinfo {author} {\bibfnamefont {M.}~\bibnamefont {Guo}}, \bibinfo {author}
  {\bibfnamefont {K.}~\bibnamefont {Li}}, \bibinfo {author} {\bibfnamefont
  {J.}~\bibnamefont {Zhang}}, \bibinfo {author} {\bibfnamefont
  {Y.}~\bibnamefont {Ou}}, \bibinfo {author} {\bibfnamefont {Y.}~\bibnamefont
  {Feng}}, \bibinfo {author} {\bibfnamefont {L.}~\bibnamefont {Wang}}, \bibinfo
  {author} {\bibfnamefont {X.}~\bibnamefont {Chen}}, \bibinfo {author}
  {\bibfnamefont {K.}~\bibnamefont {He}}, \bibinfo {author} {\bibfnamefont
  {X.}~\bibnamefont {Ma}}, \bibinfo {author} {\bibfnamefont {Q.}~\bibnamefont
  {Xue}},\ and\ \bibinfo {author} {\bibfnamefont {Y.}~\bibnamefont {Wang}},\
  }\href {https://doi.org/10.1038/ncomms5915} {\bibfield  {journal} {\bibinfo
  {journal} {Nat. Commun.}\ }\textbf {\bibinfo {volume} {5}},\ \bibinfo {pages}
  {4915} (\bibinfo {year} {2014})}\BibitemShut {NoStop}%
\bibitem [{\citenamefont {Zhao}\ \emph {et~al.}(2016)\citenamefont {Zhao},
  \citenamefont {Cheng}, \citenamefont {Pan}, \citenamefont {Zhang},
  \citenamefont {Wang}, \citenamefont {Wang}, \citenamefont {Xiu},\ and\
  \citenamefont {Song}}]{Zhao2016}%
  \BibitemOpen
  \bibfield  {author} {\bibinfo {author} {\bibfnamefont {B.}~\bibnamefont
  {Zhao}}, \bibinfo {author} {\bibfnamefont {P.}~\bibnamefont {Cheng}},
  \bibinfo {author} {\bibfnamefont {H.}~\bibnamefont {Pan}}, \bibinfo {author}
  {\bibfnamefont {S.}~\bibnamefont {Zhang}}, \bibinfo {author} {\bibfnamefont
  {B.}~\bibnamefont {Wang}}, \bibinfo {author} {\bibfnamefont {G.}~\bibnamefont
  {Wang}}, \bibinfo {author} {\bibfnamefont {F.}~\bibnamefont {Xiu}},\ and\
  \bibinfo {author} {\bibfnamefont {F.}~\bibnamefont {Song}},\ }\href
  {https://doi.org/10.1038/srep22377} {\bibfield  {journal} {\bibinfo
  {journal} {Sci. Rep.}\ }\textbf {\bibinfo {volume} {6}},\ \bibinfo {pages}
  {2237} (\bibinfo {year} {2016})}\BibitemShut {NoStop}%
\bibitem [{\citenamefont {Wei}\ \emph {et~al.}(2013)\citenamefont {Wei},
  \citenamefont {Katmis}, \citenamefont {Assaf}, \citenamefont {Steinberg},
  \citenamefont {Jarillo-Herrero}, \citenamefont {Heiman},\ and\ \citenamefont
  {Moodera}}]{Peng}%
  \BibitemOpen
  \bibfield  {author} {\bibinfo {author} {\bibfnamefont {P.}~\bibnamefont
  {Wei}}, \bibinfo {author} {\bibfnamefont {F.}~\bibnamefont {Katmis}},
  \bibinfo {author} {\bibfnamefont {B.~A.}\ \bibnamefont {Assaf}}, \bibinfo
  {author} {\bibfnamefont {H.}~\bibnamefont {Steinberg}}, \bibinfo {author}
  {\bibfnamefont {P.}~\bibnamefont {Jarillo-Herrero}}, \bibinfo {author}
  {\bibfnamefont {D.}~\bibnamefont {Heiman}},\ and\ \bibinfo {author}
  {\bibfnamefont {J.~S.}\ \bibnamefont {Moodera}},\ }\href
  {https://doi.org/10.1103/PhysRevLett.110.186807} {\bibfield  {journal}
  {\bibinfo  {journal} {Phys. Rev. Lett.}\ }\textbf {\bibinfo {volume} {110}},\
  \bibinfo {pages} {186807} (\bibinfo {year} {2013})}\BibitemShut {NoStop}%
\bibitem [{\citenamefont {Yang}\ \emph {et~al.}(2013)\citenamefont {Yang},
  \citenamefont {Dolev}, \citenamefont {Zhang}, \citenamefont {Zhao},
  \citenamefont {Fried}, \citenamefont {Schemm}, \citenamefont {Liu},
  \citenamefont {Palevski}, \citenamefont {Marshall}, \citenamefont {Risbud},\
  and\ \citenamefont {Kapitulnik}}]{Yang}%
  \BibitemOpen
  \bibfield  {author} {\bibinfo {author} {\bibfnamefont {Q.~I.}\ \bibnamefont
  {Yang}}, \bibinfo {author} {\bibfnamefont {M.}~\bibnamefont {Dolev}},
  \bibinfo {author} {\bibfnamefont {L.}~\bibnamefont {Zhang}}, \bibinfo
  {author} {\bibfnamefont {J.}~\bibnamefont {Zhao}}, \bibinfo {author}
  {\bibfnamefont {A.~D.}\ \bibnamefont {Fried}}, \bibinfo {author}
  {\bibfnamefont {E.}~\bibnamefont {Schemm}}, \bibinfo {author} {\bibfnamefont
  {M.}~\bibnamefont {Liu}}, \bibinfo {author} {\bibfnamefont {A.}~\bibnamefont
  {Palevski}}, \bibinfo {author} {\bibfnamefont {A.~F.}\ \bibnamefont
  {Marshall}}, \bibinfo {author} {\bibfnamefont {S.~H.}\ \bibnamefont
  {Risbud}},\ and\ \bibinfo {author} {\bibfnamefont {A.}~\bibnamefont
  {Kapitulnik}},\ }\href {https://doi.org/10.1103/PhysRevB.88.081407}
  {\bibfield  {journal} {\bibinfo  {journal} {Phys. Rev. B}\ }\textbf {\bibinfo
  {volume} {88}},\ \bibinfo {pages} {081407} (\bibinfo {year}
  {2013})}\BibitemShut {NoStop}%
\bibitem [{\citenamefont {Hor}\ \emph {et~al.}(2010)\citenamefont {Hor},
  \citenamefont {Roushan}, \citenamefont {Beidenkopf}, \citenamefont {Seo},
  \citenamefont {Qu}, \citenamefont {Checkelsky}, \citenamefont {Wray},
  \citenamefont {Hsieh}, \citenamefont {Xia}, \citenamefont {Xu}, \citenamefont
  {Qian}, \citenamefont {Hasan}, \citenamefont {Ong}, \citenamefont {Yazdani},\
  and\ \citenamefont {Cava}}]{Hor}%
  \BibitemOpen
  \bibfield  {author} {\bibinfo {author} {\bibfnamefont {Y.~S.}\ \bibnamefont
  {Hor}}, \bibinfo {author} {\bibfnamefont {P.}~\bibnamefont {Roushan}},
  \bibinfo {author} {\bibfnamefont {H.}~\bibnamefont {Beidenkopf}}, \bibinfo
  {author} {\bibfnamefont {J.}~\bibnamefont {Seo}}, \bibinfo {author}
  {\bibfnamefont {D.}~\bibnamefont {Qu}}, \bibinfo {author} {\bibfnamefont
  {J.~G.}\ \bibnamefont {Checkelsky}}, \bibinfo {author} {\bibfnamefont
  {L.~A.}\ \bibnamefont {Wray}}, \bibinfo {author} {\bibfnamefont
  {D.}~\bibnamefont {Hsieh}}, \bibinfo {author} {\bibfnamefont
  {Y.}~\bibnamefont {Xia}}, \bibinfo {author} {\bibfnamefont {S.-Y.}\
  \bibnamefont {Xu}}, \bibinfo {author} {\bibfnamefont {D.}~\bibnamefont
  {Qian}}, \bibinfo {author} {\bibfnamefont {M.~Z.}\ \bibnamefont {Hasan}},
  \bibinfo {author} {\bibfnamefont {N.~P.}\ \bibnamefont {Ong}}, \bibinfo
  {author} {\bibfnamefont {A.}~\bibnamefont {Yazdani}},\ and\ \bibinfo {author}
  {\bibfnamefont {R.~J.}\ \bibnamefont {Cava}},\ }\href
  {https://doi.org/10.1103/PhysRevB.81.195203} {\bibfield  {journal} {\bibinfo
  {journal} {Phys. Rev. B}\ }\textbf {\bibinfo {volume} {81}},\ \bibinfo
  {pages} {195203} (\bibinfo {year} {2010})}\BibitemShut {NoStop}%
\bibitem [{\citenamefont {Haazen}\ \emph {et~al.}(2012)\citenamefont {Haazen},
  \citenamefont {Laloe}, \citenamefont {Nummy}, \citenamefont {Swagten},
  \citenamefont {Jarillo-Herrero}, \citenamefont {Heiman},\ and\ \citenamefont
  {Moodera}}]{Haazen}%
  \BibitemOpen
  \bibfield  {author} {\bibinfo {author} {\bibfnamefont {P.~P.~J.}\
  \bibnamefont {Haazen}}, \bibinfo {author} {\bibfnamefont {J.-B.}\
  \bibnamefont {Laloe}}, \bibinfo {author} {\bibfnamefont {T.~J.}\ \bibnamefont
  {Nummy}}, \bibinfo {author} {\bibfnamefont {H.~J.~M.}\ \bibnamefont
  {Swagten}}, \bibinfo {author} {\bibfnamefont {P.}~\bibnamefont
  {Jarillo-Herrero}}, \bibinfo {author} {\bibfnamefont {D.}~\bibnamefont
  {Heiman}},\ and\ \bibinfo {author} {\bibfnamefont {J.~S.}\ \bibnamefont
  {Moodera}},\ }\href {https://doi.org/10.1063/1.3688043} {\bibfield  {journal}
  {\bibinfo  {journal} {Appl. Phys. Lett.}\ }\textbf {\bibinfo {volume}
  {100}},\ \bibinfo {pages} {082404} (\bibinfo {year} {2012})}\BibitemShut
  {NoStop}%
\bibitem [{\citenamefont {Li}\ \emph {et~al.}(2016)\citenamefont {Li},
  \citenamefont {He}, \citenamefont {Lu}, \citenamefont {Zhang}, \citenamefont
  {Liu}, \citenamefont {Ma}, \citenamefont {Fan}, \citenamefont {Shen},\ and\
  \citenamefont {Wang}}]{Li}%
  \BibitemOpen
  \bibfield  {author} {\bibinfo {author} {\bibfnamefont {H.}~\bibnamefont
  {Li}}, \bibinfo {author} {\bibfnamefont {H.}~\bibnamefont {He}}, \bibinfo
  {author} {\bibfnamefont {H.-Z.}\ \bibnamefont {Lu}}, \bibinfo {author}
  {\bibfnamefont {H.}~\bibnamefont {Zhang}}, \bibinfo {author} {\bibfnamefont
  {H.}~\bibnamefont {Liu}}, \bibinfo {author} {\bibfnamefont {R.}~\bibnamefont
  {Ma}}, \bibinfo {author} {\bibfnamefont {Z.}~\bibnamefont {Fan}}, \bibinfo
  {author} {\bibfnamefont {S.-Q.}\ \bibnamefont {Shen}},\ and\ \bibinfo
  {author} {\bibfnamefont {J.}~\bibnamefont {Wang}},\ }\href
  {https://doi.org/10.1038/ncomms10301} {\bibfield  {journal} {\bibinfo
  {journal} {Nat. Commun.}\ }\textbf {\bibinfo {volume} {7}},\ \bibinfo {pages}
  {10301} (\bibinfo {year} {2016})}\BibitemShut {NoStop}%
\bibitem [{\citenamefont {Liu}\ \emph {et~al.}(2015)\citenamefont {Liu},
  \citenamefont {Zhang}, \citenamefont {Yuan}, \citenamefont {Lei},
  \citenamefont {Wang}, \citenamefont {Di~Sante}, \citenamefont {Narayan},
  \citenamefont {He}, \citenamefont {Picozzi}, \citenamefont {Sanvito},
  \citenamefont {Che},\ and\ \citenamefont {Xiu}}]{Liu2015}%
  \BibitemOpen
  \bibfield  {author} {\bibinfo {author} {\bibfnamefont {Y.}~\bibnamefont
  {Liu}}, \bibinfo {author} {\bibfnamefont {C.}~\bibnamefont {Zhang}}, \bibinfo
  {author} {\bibfnamefont {X.}~\bibnamefont {Yuan}}, \bibinfo {author}
  {\bibfnamefont {T.}~\bibnamefont {Lei}}, \bibinfo {author} {\bibfnamefont
  {C.}~\bibnamefont {Wang}}, \bibinfo {author} {\bibfnamefont {D.}~\bibnamefont
  {Di~Sante}}, \bibinfo {author} {\bibfnamefont {A.}~\bibnamefont {Narayan}},
  \bibinfo {author} {\bibfnamefont {L.}~\bibnamefont {He}}, \bibinfo {author}
  {\bibfnamefont {S.}~\bibnamefont {Picozzi}}, \bibinfo {author} {\bibfnamefont
  {S.}~\bibnamefont {Sanvito}}, \bibinfo {author} {\bibfnamefont
  {R.}~\bibnamefont {Che}},\ and\ \bibinfo {author} {\bibfnamefont
  {F.}~\bibnamefont {Xiu}},\ }\href {https://doi.org/10.1038/am.2015.110}
  {\bibfield  {journal} {\bibinfo  {journal} {NPG Asia Materials}\ }\textbf
  {\bibinfo {volume} {7}},\ \bibinfo {pages} {e221} (\bibinfo {year}
  {2015})}\BibitemShut {NoStop}%
\bibitem [{\citenamefont {Zhang}\ \emph {et~al.}(2015)\citenamefont {Zhang},
  \citenamefont {Liu}, \citenamefont {Wang}, \citenamefont {Zhang},
  \citenamefont {Zhou}, \citenamefont {Chen}, \citenamefont {Zou},\ and\
  \citenamefont {Xiu}}]{zhang2015}%
  \BibitemOpen
  \bibfield  {author} {\bibinfo {author} {\bibfnamefont {E.}~\bibnamefont
  {Zhang}}, \bibinfo {author} {\bibfnamefont {Y.}~\bibnamefont {Liu}}, \bibinfo
  {author} {\bibfnamefont {W.}~\bibnamefont {Wang}}, \bibinfo {author}
  {\bibfnamefont {C.}~\bibnamefont {Zhang}}, \bibinfo {author} {\bibfnamefont
  {P.}~\bibnamefont {Zhou}}, \bibinfo {author} {\bibfnamefont {Z.-G.}\
  \bibnamefont {Chen}}, \bibinfo {author} {\bibfnamefont {J.}~\bibnamefont
  {Zou}},\ and\ \bibinfo {author} {\bibfnamefont {F.}~\bibnamefont {Xiu}},\
  }\href {https://doi.org/10.1021/acsnano.5b02243} {\bibfield  {journal}
  {\bibinfo  {journal} {ACS Nano}\ }\textbf {\bibinfo {volume} {9}},\ \bibinfo
  {pages} {8843–8850} (\bibinfo {year} {2015})}\BibitemShut {NoStop}%
\bibitem [{\citenamefont {Gennes}(1966)}]{DEGENNES196610}%
  \BibitemOpen
  \bibfield  {author} {\bibinfo {author} {\bibfnamefont {P.~D.}\ \bibnamefont
  {Gennes}},\ }\href
  {https://doi.org/https://doi.org/10.1016/0031-9163(66)90229-0} {\bibfield
  {journal} {\bibinfo  {journal} {Phys. Lett.}\ }\textbf {\bibinfo {volume}
  {23}},\ \bibinfo {pages} {10 } (\bibinfo {year} {1966})}\BibitemShut
  {NoStop}%
\bibitem [{\citenamefont {Hao}\ \emph {et~al.}(1991)\citenamefont {Hao},
  \citenamefont {Moodera},\ and\ \citenamefont {Meservey}}]{Hao1991}%
  \BibitemOpen
  \bibfield  {author} {\bibinfo {author} {\bibfnamefont {X.}~\bibnamefont
  {Hao}}, \bibinfo {author} {\bibfnamefont {J.~S.}\ \bibnamefont {Moodera}},\
  and\ \bibinfo {author} {\bibfnamefont {R.}~\bibnamefont {Meservey}},\ }\href
  {https://doi.org/10.1103/PhysRevLett.67.1342} {\bibfield  {journal} {\bibinfo
   {journal} {Phys. Rev. Lett.}\ }\textbf {\bibinfo {volume} {67}},\ \bibinfo
  {pages} {1342} (\bibinfo {year} {1991})}\BibitemShut {NoStop}%
\bibitem [{\citenamefont {Son}\ and\ \citenamefont {Spivak}(2013)}]{Son2013}%
  \BibitemOpen
  \bibfield  {author} {\bibinfo {author} {\bibfnamefont {D.~T.}\ \bibnamefont
  {Son}}\ and\ \bibinfo {author} {\bibfnamefont {B.~Z.}\ \bibnamefont
  {Spivak}},\ }\href {https://doi.org/10.1103/PhysRevB.88.104412} {\bibfield
  {journal} {\bibinfo  {journal} {Phys. Rev. B}\ }\textbf {\bibinfo {volume}
  {88}},\ \bibinfo {pages} {104412} (\bibinfo {year} {2013})}\BibitemShut
  {NoStop}%
\bibitem [{\citenamefont {Burkov}(2014)}]{Burkov2014}%
  \BibitemOpen
  \bibfield  {author} {\bibinfo {author} {\bibfnamefont {A.~A.}\ \bibnamefont
  {Burkov}},\ }\href {https://doi.org/10.1103/PhysRevLett.113.247203}
  {\bibfield  {journal} {\bibinfo  {journal} {Phys. Rev. Lett.}\ }\textbf
  {\bibinfo {volume} {113}},\ \bibinfo {pages} {247203} (\bibinfo {year}
  {2014})}\BibitemShut {NoStop}%
\bibitem [{\citenamefont {Li}\ \emph {et~al.}(2015)\citenamefont {Li},
  \citenamefont {Wang}, \citenamefont {Liu}, \citenamefont {Wang},
  \citenamefont {Liao},\ and\ \citenamefont {Yu}}]{Li2015}%
  \BibitemOpen
  \bibfield  {author} {\bibinfo {author} {\bibfnamefont {C.-Z.}\ \bibnamefont
  {Li}}, \bibinfo {author} {\bibfnamefont {L.-X.}\ \bibnamefont {Wang}},
  \bibinfo {author} {\bibfnamefont {H.}~\bibnamefont {Liu}}, \bibinfo {author}
  {\bibfnamefont {J.}~\bibnamefont {Wang}}, \bibinfo {author} {\bibfnamefont
  {Z.-M.}\ \bibnamefont {Liao}},\ and\ \bibinfo {author} {\bibfnamefont
  {D.-P.}\ \bibnamefont {Yu}},\ }\href {https://doi.org/10.1038/ncomms10137}
  {\bibfield  {journal} {\bibinfo  {journal} {Nat. Commun.}\ }\textbf {\bibinfo
  {volume} {6}},\ \bibinfo {pages} {10137} (\bibinfo {year}
  {2015})}\BibitemShut {NoStop}%
\bibitem [{\citenamefont {Zhang}\ \emph {et~al.}(2011)\citenamefont {Zhang},
  \citenamefont {Richard}, \citenamefont {Qian}, \citenamefont {Xu},
  \citenamefont {Dai},\ and\ \citenamefont {Ding}}]{Zhang2011}%
  \BibitemOpen
  \bibfield  {author} {\bibinfo {author} {\bibfnamefont {P.}~\bibnamefont
  {Zhang}}, \bibinfo {author} {\bibfnamefont {P.}~\bibnamefont {Richard}},
  \bibinfo {author} {\bibfnamefont {T.}~\bibnamefont {Qian}}, \bibinfo {author}
  {\bibfnamefont {Y.-M.}\ \bibnamefont {Xu}}, \bibinfo {author} {\bibfnamefont
  {X.}~\bibnamefont {Dai}},\ and\ \bibinfo {author} {\bibfnamefont
  {H.}~\bibnamefont {Ding}},\ }\href {https://doi.org/10.1063/1.3585113}
  {\bibinfo {title} {A precise method for visualizing dispersive features in
  image plots}} (\bibinfo {year} {2011})\BibitemShut {NoStop}%
\bibitem [{\citenamefont {Kandala}\ \emph {et~al.}(2013)\citenamefont
  {Kandala}, \citenamefont {Richardella}, \citenamefont {Rench}, \citenamefont
  {Zhang}, \citenamefont {Flanagan},\ and\ \citenamefont
  {Samarth}}]{Kandala_2013}%
  \BibitemOpen
  \bibfield  {author} {\bibinfo {author} {\bibfnamefont {A.}~\bibnamefont
  {Kandala}}, \bibinfo {author} {\bibfnamefont {A.}~\bibnamefont
  {Richardella}}, \bibinfo {author} {\bibfnamefont {D.~W.}\ \bibnamefont
  {Rench}}, \bibinfo {author} {\bibfnamefont {D.~M.}\ \bibnamefont {Zhang}},
  \bibinfo {author} {\bibfnamefont {T.~C.}\ \bibnamefont {Flanagan}},\ and\
  \bibinfo {author} {\bibfnamefont {N.}~\bibnamefont {Samarth}},\ }\href
  {https://doi.org/10.1063/1.4831987} {\bibfield  {journal} {\bibinfo
  {journal} {Appl. Phys. Lett.}\ }\textbf {\bibinfo {volume} {103}},\ \bibinfo
  {pages} {202409} (\bibinfo {year} {2013})}\BibitemShut {NoStop}%
\bibitem [{\citenamefont {Yang}\ \emph {et~al.}(2019)\citenamefont {Yang},
  \citenamefont {Fanchiang}, \citenamefont {Chen}, \citenamefont {Tseng},
  \citenamefont {Liu}, \citenamefont {Guo}, \citenamefont {Hong}, \citenamefont
  {Lee},\ and\ \citenamefont {Kwo}}]{Yang_PhysRevB.100.045138}%
  \BibitemOpen
  \bibfield  {author} {\bibinfo {author} {\bibfnamefont {S.~R.}\ \bibnamefont
  {Yang}}, \bibinfo {author} {\bibfnamefont {Y.~T.}\ \bibnamefont {Fanchiang}},
  \bibinfo {author} {\bibfnamefont {C.~C.}\ \bibnamefont {Chen}}, \bibinfo
  {author} {\bibfnamefont {C.~C.}\ \bibnamefont {Tseng}}, \bibinfo {author}
  {\bibfnamefont {Y.~C.}\ \bibnamefont {Liu}}, \bibinfo {author} {\bibfnamefont
  {M.~X.}\ \bibnamefont {Guo}}, \bibinfo {author} {\bibfnamefont
  {M.}~\bibnamefont {Hong}}, \bibinfo {author} {\bibfnamefont {S.~F.}\
  \bibnamefont {Lee}},\ and\ \bibinfo {author} {\bibfnamefont {J.}~\bibnamefont
  {Kwo}},\ }\href {https://doi.org/10.1103/PhysRevB.100.045138} {\bibfield
  {journal} {\bibinfo  {journal} {Phys. Rev. B}\ }\textbf {\bibinfo {volume}
  {100}},\ \bibinfo {pages} {045138} (\bibinfo {year} {2019})}\BibitemShut
  {NoStop}%
\bibitem [{\citenamefont {Yakovleva}\ \emph {et~al.}(2015)\citenamefont
  {Yakovleva}, \citenamefont {Oveshnikov}, \citenamefont {Kochura},
  \citenamefont {Lisunov}, \citenamefont {Lahderanta},\ and\ \citenamefont
  {Aronzon}}]{yakovleva2015anomalous}%
  \BibitemOpen
  \bibfield  {author} {\bibinfo {author} {\bibfnamefont {E.}~\bibnamefont
  {Yakovleva}}, \bibinfo {author} {\bibfnamefont {L.~N.}\ \bibnamefont
  {Oveshnikov}}, \bibinfo {author} {\bibfnamefont {A.}~\bibnamefont {Kochura}},
  \bibinfo {author} {\bibfnamefont {K.}~\bibnamefont {Lisunov}}, \bibinfo
  {author} {\bibfnamefont {E.}~\bibnamefont {Lahderanta}},\ and\ \bibinfo
  {author} {\bibfnamefont {B.~A.}\ \bibnamefont {Aronzon}},\ }\href
  {https://link.springer.com/article/10.1134/S0021364015020149} {\bibfield
  {journal} {\bibinfo  {journal} {JETP Letters}\ }\textbf {\bibinfo {volume}
  {101}},\ \bibinfo {pages} {130} (\bibinfo {year} {2015})}\BibitemShut
  {NoStop}%
\end{thebibliography}
\providecommand{\noopsort}[1]{}\providecommand{\singleletter}[1]{#1}%

\end{document}